\documentclass[12pt]{article}

\usepackage{amsmath,amssymb}
\usepackage{slashed}
\usepackage{xcolor} 
\usepackage{graphicx}
\usepackage{url}
\usepackage{cite}

\usepackage{verbatim}
   \allowdisplaybreaks
\makeatletter
\@addtoreset{equation}{section}
\makeatother

\graphicspath{{./Figures/}}

\newcommand{\be}{\begin{equation}} 
\newcommand{\ee}{\end{equation}} 
\newcommand{\bea}{\begin{eqnarray}}  
\newcommand{\eea}{\end{eqnarray}}
\newcommand{\bs}{\begin{split}} 
\newcommand{\es}{\end{split}}

\newcommand{\tr}{\operatorname{tr}}

\begin{document}
\thispagestyle{empty}

\begin{center}
\hfill CERN-PH-TH-2015-124
\begin{center}

\vspace{.5cm}

{\Large\sc GMSB with Light Stops}

\end{center}
\vspace{1.cm}

\textbf{ Antonio Delgado$^{\,a,b}$, Mateo Garcia-Pepin$^{\,c}$,
and Mariano Quiros$^{\,d}$}\\

\vspace{.1cm}
${}^a\!\!$ {\em Department of Physics, 225 Nieuwland Science Hall, University of Notre Dame,\\ Notre Dame, IN 46556, USA}

\vspace{.1cm}
${}^b\!\!$ {\em Theory Division, Physics Department CERN,\\ CH-1211 Geneva 23, Switzerland}

\vspace{.1cm}
${}^c\!\!$ {\em {Institut de F\'isica d'Altes Energies, Universitat Aut{\`o}noma de Barcelona\\
08193 Bellaterra, Barcelona, Spain}}

\vspace{.1cm}
${}^d\!\!$ {\em {Instituci\'o Catalana de Recerca i Estudis  
Avan\c{c}ats (ICREA) and\\ Institut de F\'isica d'Altes Energies, Universitat Aut{\`o}noma de Barcelona\\
08193 Bellaterra, Barcelona, Spain}}

\end{center}

\vspace{0.8cm}

\centerline{\bf Abstract}
\vspace{2 mm}
\begin{quote}\small
Gauge mediated supersymmetry breaking (GMSB) is an elegant mechanism to transmit supersymmetry breaking from the hidden to the MSSM observable sector, which solves the supersymmetric flavor problem. However the smallness of the generated stop mixing requires superheavy stops to reproduce the experimental value of the Higgs mass.  
A possible way out is to extend the MSSM Higgs sector with singlets and/or triplets providing extra tree-level corrections to the Higgs mass. Singlets will not get any soft mass from GMSB and triplets will contribute to the $\rho$ parameter which could be an issue. In this paper we explore the second possibility by introducing extra supersymmetric triplets with hypercharges $Y=(0,\pm 1)$, with a tree-level custodial $SU(2)_L\otimes SU(2)_R$ global symmetry in the Higgs sector protecting the $\rho$ parameter: a supersymmetric generalization of the Georgi-Machacek model, dubbed as supersymmetric custodial triplet model (SCTM). The renormalization group running from the messenger to the electroweak scale mildly breaks the custodial symmetry. We will present realistic low-scale scenarios (with the NLSP being a Bino-like neutralino or the right-handed stau) based on general (non-minimal) gauge mediation and consistent with all present experimental data. Their main features are: i) Light ($\sim  1$ TeV) stops; ii) Exotic couplings ($H^\pm W^\mp Z$ and $H^{\pm\pm} W^\mp W^\mp$) absent in the MSSM and proportional to the triplets VEV, $v_\Delta$; and, iii) A possible (measurable)  universality breaking of the Higgs couplings $\lambda_{WZ}=r_{WW}/r_{ZZ}\neq 1$.

\end{quote}

\vfill

\newpage
\section{Introduction}
\label{Introduction} 

Among a few other possibilities, supersymmetry remains as the simplest, perturbative solution to the Higgs hierarchy problem of the Standard Model (SM). Particularly interesting is the minimal supersymmetric extension of the SM, dubbed as MSSM, on which most of the experimental detection efforts are concentrated at the Large Hadron Collider (LHC). In spite of its simplicity, the mechanism of supersymmetry breaking in the observable sector is an unsettled issue. Supersymmetry is usually assumed to be broken in a hidden sector and communicated to the observable sector. Depending on the mediation mechanism the supersymmetric theory can introduce flavor violating interactions spoiling its phenomenological viability, a problem known as the \textit{supersymmetric flavor problem}. This problem is automatically solved by gauge mediated supersymmetry breaking (GMSB) models~\cite{Giudice:1998bp}, as the gauge interactions are flavor diagonal, provided that the scale of messengers is low enough so that the gravitational contributions can be neglected. 

A main feature of GMSB in the MSSM is that the predicted value of the stop mixing parameter $A_t$ is very small at the messenger scale $\mathcal M$, as it comes from two-loop diagrams. As a consequence, the discovery of the Higgs boson with a mass around 125 GeV, has somewhat jeopardized GMSB theories for the MSSM since, in order to reproduce the Higgs mass, stops heavier than 5 TeV are required~\cite{Delgado:2013gza,Draper:2011aa,Draper:2013oza}. This in turn would reintroduce a little hierarchy problem and stops would be very far away from the LHC reach. 

Two options appear to tackle this problem in GMSB theories. One option is increasing the value of the radiative contributions to the Higgs mass, either by generating large values of the mixing parameter $A_t$, or by enlarging the MSSM with heavy vector-like fermions strongly coupled to the Higgs sector~\cite{Nickel:2015dna}. 
In particular, generating large values of $A_t$ can be done by introducing direct messenger-MSSM superpotential couplings~\cite{Craig:2013wga,Evans:2013kxa,Evans:2015swa,Byakti:2013ti,Kang:2012ra,Evans:2011bea,Calibbi:2013mka,Calibbi:2014yha,Chacko:2001km,Shadmi:2011hs}.  These models, dubbed extended GMSB, do not necessarily lead to minimal flavor violation (MFV) and the flavor constraints require a special flavor texture.  In both scenarios the fine-tuning is considerably reduced with respect to that in the MSSM with GMSB. The second option, without enlarging the SM gauge group, is increasing the value of the Higgs mass by means of a tree-level $F$-term from an extended MSSM Higgs sector. This second option will be considered in this paper.

The MSSM extensions which can increase the Higgs mass by a tree-level $F$-term are limited to fields in the superpotential which can couple at the renormailzable level to the MSSM Higgs sector $H_{1,2}$~\footnote{In our notation $H_2$ gives a mass to the top quark and $H_1$ to the bottom quark and charged lepton.}: they are a singlet $S$ and/or triplets with hypercharge $Y=(0,\pm 1$), $\Sigma_{0,\,\pm 1}$. Any of the above extra Higgses would add (depending on the value of $\tan\beta$) an extra tree-level contribution to the Higgs mass. Following our previous philosophy we can exclude the presence of the singlet, as it does not get any mass from the GMSB unless: \textit{i)} We enlarge the gauge group such that $S$ transforms as a non-trivial representation of the enlarged gauge group, or; \textit{ii)} We consider an extended GMSB model with direct superpotential messenger-MSSM couplings~\cite{Delgado:2007rz}, which could result again in flavor constraints. The only surviving possibility is then adding the triplets $\Sigma_{0,\,\pm 1}$.

Introducing only $\Sigma_{0}$ or $\Sigma_{\pm 1}$ has a general problem as the neutral component of the triplets will acquire a vacuum expectation value (VEV) $v_\Delta$, which will spoil the $\rho=1$ relationship unless $v_\Delta$ is small enough, which requires a large soft mass for the triplet. Since the contribution to each mass is tied by the the gauge structure of the theory, it will be impossible for gauge mediation to generate large $SU(2)_L$ triplet masses while keeping the rest of the spectrum light. Therefore trying to solve the $\rho=1$ problem in this way would recreate a strong naturalness (little hierarchy) problem. The way out is using the whole set $\Sigma_{0,\,\pm 1}$ and 
providing the theory with a global $SU(2)_L\otimes SU(2)_R$ symmetry, spontaneously  broken to the custodial $SU(2)_V$ symmetry after electroweak (EW) breaking. This kind of models were first introduced in the context of nonsupersymmetric theories by Georgi and Machacek (GM)~\cite{Georgi:1985nv}, generalized to supersymmetric theories in Ref.~\cite{Cort:2013foa} and further explored in~\cite{Garcia-Pepin:2014yfa}. It is dubbed supersymmetric custodial triplet model (SCTM). 

The SCTM model makes use of the custodial symmetry to solve the $\rho$ problem of theories with triplets. Custodial boundary conditions for the Higgs sector are required although custodial symmetry is spoiled by radiative corrections, proportional to the hypercharge and top Yukawa couplings. Therefore the renormalization group equation (RGE) running departs from the custodial symmetry conditions. One can then allow for some departure from the $\rho=1$ custodial solution but not too much: this can be fulfilled in a GMSB mechanism provided that the messenger scale $\mathcal M$ is low enough (a natural condition in GMSB models). Moreover GMSB provides custodial boundary values to the Higgs sector, except for the contribution of the hypercharge coupling which will break explicitly custodial invariance. As we will see, this explicit breaking will not change the main features nor the phenomenology of the model. Of course this model is able to raise the tree level Higgs mass through new $F$-term contributions and fit the $\sim 125$ GeV measurement without the need of super-heavy stops. At the same time it generates large triplet VEVs that can participate in the EW breaking up to a $\sim 15\%$ order.

In this paper we will define a non-minimal gauge mediation mechanism which will provide a soft spectrum for the SCTM making it consistent with all electroweak and LHC data, and thus alleviating the tension between the Higgs mass, light stops and the supersymmetric flavor problem. The rest of the paper is organized as follows. In Sec.~\ref{sec:model} we will describe the model and its particular vacuum structure. The implementation of a gauge mediated mechanism of supersymmetry breaking is discussed in Sec.~\ref{sec:gauge} and typical benchmark scenarios are proposed in Sec.~{\ref{sec:benchmark}. A study on the phenomenology and collider features is performed in Sec.~\ref{sec:pheno}. We finally discuss our conclusions in Sec.~\ref{sec:discussion}.

\section{The Model}
\label{sec:model}
At the scale $\mathcal M$ at which supersymmetry breaking is transmitted to the observable sector we assume the supersymmetric theory to be invariant under $SU(2)_L\otimes SU(2)_R$ broken only by Yukawa and hypercharge interactions. We add to the MSSM Higgs sector $H_1$ and $H_2$, with respective hypercharges $Y=(-1/2,\,1/2)$ 
 \be
   H_1=\left( \begin{array}{c}H_1^0\\ H_1^-\end{array}\right),\quad
   H_2=\left( \begin{array}{c}H_2^+\\ H_2^0\end{array}\right)
   \ee
three $SU(2)_L$ triplets, $\Sigma_{-1}$, $\Sigma_0$ and $\Sigma_1$ with hypercharges  $Y=(-1,\, 0,\, 1)$, which we represent by two dimensional matrices as 
 \be
 \Sigma_{-1}=\left(\begin{array}{cc} \frac{\chi^-}{\sqrt{2}} & \chi^0\\\chi^{--}& -\frac{\chi^-}{\sqrt{2}}
 \end{array}
 \right),\quad  \Sigma_{0}=\left(\begin{array}{cc} \frac{\phi^0}{\sqrt{2}} & \phi^+\\ \phi^{-}& -\frac{\phi^0}{\sqrt{2}}
 \end{array}
 \right),\quad  \Sigma_{1}=\left(\begin{array}{cc} \frac{\psi^+}{\sqrt{2}} & \psi^{++}\\\psi^{0}& -\frac{\psi^+}{\sqrt{2}}
 \end{array}
 \right)\ .
 \ee
where $Q=T_{3L}+Y$. They are organized under $SU(2)_L\otimes SU(2)_R$ as $\bar H=(\textbf{2},\bar {\textbf{2}})$, and $\bar\Delta=(\textbf{3},\bar{\textbf{3}})$ where
\be
\bar H=\left( \begin{array}{c}H_1\\ H_2\end{array}\right),\quad
\bar \Delta=\left(\begin{array}{cc} -\frac{\Sigma_0}{\sqrt{2}} & -\Sigma_{-1}\\ -\Sigma_{1}& \frac{\Sigma_0}{\sqrt{2}}\end{array}\right)
\ee
and $\bar T_{3R}=-T_{3R}=Y$.   The invariant products for doublets $A\cdot B\equiv A^a\epsilon_{ab}B^b$  and anti-doublets $\bar A\cdot \bar B\equiv\bar A_a\epsilon^{ab}\bar B_c$ are defined by $\epsilon_{21}=\epsilon^{12}=1$. The $SU(2)_L\otimes SU(2)_R$ invariant superpotential is defined as
\be
W_0=\lambda \bar H\cdot \bar\Delta\bar H+\frac{\lambda_3}{3}\tr\bar\Delta^3+\frac{\mu}{2}\bar H\cdot\bar H+\frac{\mu_\Delta}{2}\tr \bar\Delta^2+ h_t\,\overline Q_L\cdot H_2 t_R + h_b\,\overline Q_L\cdot H_1 b_R
\label{W0}
\ee
Gauge mediated supersymmetry breaking will generate masses at the messenger scale $\mathcal M$ for all scalars, as we will describe in detail in the next section. As we will see the mass spectrum of the Higgs scalars at the scale $\mathcal M$ is  $SU(2)_L\otimes SU(2)_R$ invariant except for contributions proportional to the $U(1)_Y$ gauge coupling which will moderately spoil the $SU(2)_L\otimes SU(2)_R$ structure of the squared mass of the triplet $\bar\Delta$. However, this violation is similar to the violation of the custodial symmetry induced by the hypercharge coupling in the RG running and does not spoil the main phenomenological features of the model.

Due to the presence of $SU(2)_L\otimes SU(2)_R$ breaking by $U(1)_Y$ and Yukawa interactions, the RGE running will split the $SU(2)_L\otimes SU(2)_R$ invariant operators into $SU(2)_L$ ones. The most general superpotential can then be written as
\be
\begin{aligned}
W &= - \lambda_aH_1\cdot\Sigma_1H_1+\lambda_b H_2\cdot\Sigma_{-1}H_2+\sqrt{2}\lambda_cH_1\cdot\Sigma_0H_2+\sqrt{2}\lambda_3\text{tr}\,\Sigma_1\Sigma_0\Sigma_{-1} \\ & 
 -  \mu H_1\cdot H_2 + \frac{\mu_{\Delta_a}}{2}\text{tr}\,\Sigma_0^2 + \mu_{\Delta_b}\text{tr}\,\Sigma_1\Sigma_{-1}+ h_t\,\overline Q_L\cdot H_2 t_R + h_b\,\overline Q_L\cdot H_1 b_R
\end{aligned}
\ee
where the $SU(2)_L\otimes SU(2)_R$ invariant situation is recovered when $\lambda_a=\lambda_b=\lambda_c \equiv \lambda$ and $\mu_{\Delta_a}=\mu_{\Delta_b}\equiv \mu_\Delta$. The total potential is then
$V=V_F+V_D+V_{\text{SOFT}}$, where
\be
\begin{aligned}
V_{\text{SOFT}}&=m_{H_1}^2H_1^\dagger H_1+m_{H_2}^2H_2^\dagger H_2+m_{\Sigma_0}^2\Sigma_0^\dagger \Sigma_0+m_{\Sigma_1}^2\Sigma_1^\dagger \Sigma_1+m_{\Sigma_{-1}}^2\Sigma_{-1}^\dagger \Sigma_{-1} -  m_3^2 H_1\cdot H_2 \\ +&\left\{\frac{B_{\Delta_a}}{2}\text{tr}\Sigma_0^2+B_{\Delta_b}\text{tr}\Sigma_1\Sigma_{-1}-A_{\lambda_a}H_1\cdot\Sigma_1H_1+A_{\lambda_b}H_2\cdot\Sigma_{-2}H_2\right. \\ 
&\left.+\sqrt{2}A_{\lambda_c}H_1\cdot\Sigma_0H_2+\sqrt{2}A_{\lambda_3}\text{tr}\,\Sigma_1\Sigma_0\Sigma_{-1}+ a_t\,\tilde Q_L\cdot H_2 \tilde t_R+ a_b\,\tilde Q_L\cdot H_1\tilde{b}_R+h.c.\right\}
\end{aligned}
\ee
and the $SU(2)_L\otimes SU(2)_R$ conditions in the supersymmetry breaking sector would be given by: $m_{H_1}=m_{H_2}\equiv m_H$, $m_{\Sigma_0}=m_{\Sigma_1}=m_{\Sigma_{-1}}\equiv m_{\Sigma}$, $B_{\Delta_a}=B_{\Delta_b}\equiv B_{\Delta}$, $A_{\lambda_a}=A_{\lambda_b}=A_{\lambda_c}\equiv A_\lambda$.
We now expand the neutral components of the fields in a totally general way as in Ref.~\cite{Garcia-Pepin:2014yfa}
$X =\frac{1}{\sqrt{2}}\left( v_X + X_R +\imath X_I \right)$, where $X=H^0_1,H^0_2,\phi^0,\chi^0,\psi^0$,
and we parametrize the departure from custodial symmetry through three angles as 
\begin{eqnarray}
v_1&=& \sqrt{2}\cos\beta v_H,~~v_2=\sqrt{2}\sin\beta v_H, \nonumber\\
v_\psi &=& 2\cos\theta_1\cos\theta_0 v_\Delta,~~v_\chi=2\sin\theta_1\cos\theta_0 v_\Delta,\nonumber\\
v_\phi &=& \sqrt{2} \sin\theta_0 v_\Delta .
\label{eq:vacio}
\end{eqnarray}
The parametrization preserves the relation
\be
\label{vev246}
v^2 \equiv (246\, \text{GeV})^2= 2v_H^2+8v_{\Delta}^2\, ,
\ee
and we recover the $SU(2)_V$ invariant vacuum when $\tan{\beta}=\tan{\theta_0}=\tan{\theta_1}=1$ 
\be
\label{cusvacuum}
 v_1=v_2\equiv v_H\quad \textrm{and}\quad v_\psi=v_\chi=v_\phi\equiv v_\Delta\ .
\ee 

We can parametrize the contribution to the deviation from $\rho=1$ from these new extra states in the following form
\be
\label{rho}
\Delta\rho=\frac{2( 2v_\phi^2-v_\psi^2-v_\chi^2)}{v_1^2+v_2^2+4(v_\chi^2+v_\psi^2)} =-4\frac{\cos 2\theta_0 v_\Delta^2}{v_H^2+8\cos^2\theta_0v_\Delta^2}
\ee
where we define $\rho\equiv 1+\Delta\rho$. One can see from this equation that, for $v_\Delta\neq 0$, a necessary and sufficient condition for the tree level condition $\rho=1$ is $\tan{\theta_0}=1$. This direction of the vacuum (which contains the custodial point $\tan{\beta}=\tan{\theta_0}=\tan{\theta_1}=1$~\footnote{Notice that the custodial condition $\tan{\beta}=\tan{\theta_0}=\tan{\theta_1}=1$ is certainly sufficient but not necessary for the tree level condition $\rho=1$. This case is reminiscent of the MSSM where the custodial condition $\tan\beta=1$ is  not necessary for the fulfillment of the tree-level condition $\rho=1$.}), will be critical for the study of the viability of the model. As it was already pointed out in~\cite{Garcia-Pepin:2014yfa} the requirement that the superpotential is a holomorphic function in the fields opens up this direction, making the model viable from a UV perspective as opposed to the non-SUSY versions where the custodial symmetry is required by the tree-level condition $\rho=1$~\footnote{Note that in the case of the non-SUSY GM model it turns out that $\tan\theta_1\equiv 1$ identically so the condition $\tan\theta_0=1$ is equivalent to the custodial symmetry in the triplet sector.}. 

If we want to explore the model at the EW scale we need to solve the Equations of Minimum (EoM) ensuring correct EW breaking. Five neutral scalar fields will generate five minimization conditions that will fix five parameters. Since we are working on a top down approach, where we will run down from the messenger scale $\mathcal M$ to the EW scale, we will need to keep consistency between the boundary conditions and the EoMs. As the parameters $m_3^2$ and $B_{\Delta_{a,b}}$ have their RGEs decoupled from the rest, we can consistently fix two of them, as e.g.~$m_3^2$ and $B_{\Delta_a}$ at the weak scale. The value of $B_{\Delta_b}$ at the weak scale will be consistently fixed in agreement with its EoM by choosing at the messenger scale $\mathcal M$ a custodial parameter $B_\Delta$ satisfying the boundary condition $B_{\Delta_a}(\mathcal M)=B_{\Delta_b}(\mathcal M)\equiv B_{\Delta}$~\footnote{We expect the same physics responsible for generating the effective behaviour that we describe in this paper to produce the correct values of $m_3^2$, $B_{\Delta}$ at the messenger scale $\mathcal M$.}. The other three EoM self consistently determine the values of the custodial breaking angles ($\tan{\beta},\tan{\theta_0},\tan{\theta_1}$) which are then a prediction of the EoMs for a given value of $v_\Delta$.

The EoMs are just criticality conditions as they do not tell us whether we are really exploring a minimum of the potential, and much less if this minimum is the absolute one. The minimum condition will be provided by the absence of tachyonic states in the scalar spectrum. Moreover each minimum we find is likely the deepest one since it consists on a smooth deformation of an $SU(2)_V$ preserving minimum where the D-terms vanish, therefore with minimized energy.

\section{Gauge Mediation in the SCTM}
\label{sec:gauge}

In the minimal realization of gauge mediation (MGM) the messenger fields transform under $\bf{r}$ and $\bar{\bf{r}}$ representations of $SU(5)$ and feel the breaking of supersymmetry through the superpotential,
$
W=\lambda^{ij} X\Phi_i\bar{\Phi}_j
$,
where $X$ is an spurion field that parametrizes the breaking of supersymmetry in the secluded sector. As MGM provides a very rigid framework to encompass low energy phenomenology we will consider a particular model of general gauge mediation~\cite{Meade:2008wd} (GGM) where there is more flexibility to accommodate the supersymmetric mass spectrum of the SCTM. We will consider a model where messengers transform only under one of the SM gauge groups $SU(3)\otimes SU(2)_L\otimes U(1)_Y$ and will choose (non-exotic) representations which are contained in $SU(5)$. In particular, to transmit supersymmetry breaking to the observable sector, we choose the messenger representations~\footnote{$\Phi_8$ and $\Phi_3$ where already used as messengers in~\cite{Han:1998pa}.}
\be
\Phi_{8}=(\mathbf{8},\mathbf{1})_0, \quad \Phi_3=(\mathbf{1},\mathbf{3})_0\quad \text{and} \quad \left[\Phi_1=(\mathbf{1},\mathbf{1})_1,\,\bar\Phi_1=(\mathbf{1},\mathbf{1})_{-1}\right]\ .
\ee

According with GGM we will explore the more general case where the messengers have independent mass terms instead of getting all their mass from the spurion superfield. For simplicity, we also consider that the scalar component of $X$ does not acquire a VEV~\footnote{In fact we are assuming that $\langle X\rangle \ll \mathcal M_A$, $A=8,3,1$.}, thus $\langle X \rangle=\theta^2F$.
\be
W=\left(\tilde{\lambda}_8^{ij}X+\mathcal M_8^{ij}\right)\Phi_{8i}\Phi_{8j}+\left(\tilde{\lambda}_3^{ij}X+\mathcal M_3^{ij}\right)\Phi_{3i}\Phi_{3j}+\left(\tilde{\lambda}_1^{ij}X+\mathcal M_1^{ij}\right)\bar\Phi_{1i}\Phi_{1j}\ee

We now impose an $O(n_8)\otimes O(n_3)\otimes O(n_1)$ global symmetry in the superpotential, where $n_8$, $n_3$ and $n_1$ are the of number of copies of each messenger respectively~\footnote{In the case of $n_1$, it is the number of pairs $(\Phi_1,\tilde{\Phi}_1)$ due to anomaly cancelation.}. Due to this symmetry, the dot product is the only invariant that can be built, thus ensuring the diagonal form of $\tilde{\lambda}_A^{ij}$ $(\equiv \delta^{ij}\tilde{\lambda}_A)$ and $\mathcal M_A^{ij}$ $(\equiv \delta^{ij}\mathcal M_A)$ in the mass basis. Via messenger parity, this symmetry prevents dangerous one-loop contributions to the masses of sleptons~\cite{Dvali:1996cu,Dimopoulos:1996ig}. Moreover for simplicity we will consider a common messenger scale so that we will assume $\mathcal M_A\equiv \mathcal M$ ($A=8,3,1$).

Within this setup and with $\Lambda_8\equiv \tilde{\lambda}_8\Lambda$, $\Lambda_3\equiv \tilde{\lambda}_3\Lambda$ and $\Lambda_1\equiv \tilde{\lambda}_1\Lambda$ ($\Lambda\equiv F/\mathcal M$) the gaugino masses at the messenger scale are,
\be
\begin{aligned}
M_3&=\frac{\alpha_3(\mathcal M)}{4\pi}3n_8g( \Lambda_8/\mathcal M)\Lambda_8\, , \\
M_2&=\frac{\alpha_2(\mathcal M)}{4\pi}2n_3g( \Lambda_3/\mathcal M)\Lambda_3\, , \\
M_1&=\frac{\alpha_1(\mathcal M)}{4\pi}\frac{6}{5}n_1g(\Lambda_1/\mathcal M)\Lambda_1\, ,
\label{gauginos}
\end{aligned}
\ee
where we are using $SU(5)$ normalization for the $U(1)$. For sfermions,
\be
\begin{aligned}
m_{\tilde{f}}^2&=2[C_3^f\left(\frac{\alpha_3(\mathcal M)}{4\pi}\right)^23n_8f( \Lambda_8/\mathcal M)\Lambda_8^2+C_2^f\left(\frac{\alpha_2(\mathcal M)}{4\pi}\right)^22n_3f( \Lambda_3/\mathcal M)\Lambda_3^2 \\ &+C_1^f\left(\frac{\alpha_1(\mathcal M)}{4\pi}\right)^2\frac{1}{2}\left(\frac{6}{5}\right)^2n_1f(\Lambda_1/\mathcal M)\Lambda_1^2   ]\, .
\label{sfermions}
\end{aligned}
\ee
Where $C_a^f$ is the quadratic Casimir of the sfermion $\tilde f$~\footnote{It is equal to $\frac{N^2-1}{2N}$ for the fundamental $\textbf{N}$ representation of $SU(N)$ and, in our notation $C_1^f=Y_f^2$, where $Y_f$ is the SM hypercharge of $\tilde f $.}. The functions $g(x)$ and $f(x)$ come from two loop exact results and were first computed in Refs.~\cite{Dimopoulos:1996gy,Martin:1996zb} as 
\be
\begin{aligned}
g(x)&=\frac{1}{x^2}\left[(1+x)\log (1+x)\right] + (x\rightarrow -x) \\
f(x)&=\frac{1+x}{x^2}\left[\log (1+x)  - 2\text{Li}_2\left(\frac{x}{1+x}\right) +\frac{1}{2}\text{Li}_2\left(\frac{2x}{1+x}\right)\right] + (x\rightarrow -x)\ .
\end{aligned}
\ee
They become relevant for small values of $\mathcal M$, as it is our case.

As showed in Eqs.~({\ref{gauginos}) and (\ref{sfermions}) an unusual messenger sector will modify the boundary conditions at the messenger scale with respect to the minimal scenario. For instance, assuming $g(x_i)\simeq 1$ we can write, at one loop, an RGE invariant gaugino mass relation which will be different from the minimal case $M_1(\mathcal M)/\alpha_1(\mathcal M)=M_2(\mathcal M)/\alpha_2(\mathcal M)=M_3(\mathcal M)/\alpha_3(\mathcal M)$. In particular
\be
\label{eq:gauginomass}
\frac{M_1(\mathcal M)}{\alpha_1(\mathcal M)}:\frac{M_2(\mathcal M)}{\alpha_2(\mathcal M)}:\frac{M_3(\mathcal M)}{\alpha_3(\mathcal M)}=\frac{6}{5}n_1\tilde{\lambda}_1:2n_3\tilde{\lambda}_3:3n_8\tilde{\lambda}_8\,.
\ee 
This shows that, besides $\mathcal M$ and $\sqrt{F}$, the boundary conditions depend on the two sets of parameters: $(n_8,n_3,n_1)$ and $(\tilde{\lambda}_8,\tilde{\lambda}_3,\tilde{\lambda}_1)$. As a result of this, once the superpotential parameters $\mathcal M$ and $\sqrt{F}$ are fixed, the low energy features of the theory will be determined by our choice of $n_A$ and $\tilde{\lambda}_A$.

\section{Benchmark scenarios}
\label{sec:benchmark}
  
As we outlined in Sec.~\ref{Introduction} the main goal of this work is to achieve light stop masses within the context of gauge mediation. Due to the strongest color contribution, if gluinos are heavier than stops they will raise the stop masses through the RGE running, making their boundary condition at the messenger scale unimportant. In a gauge mediated context we can generally say that the heavier the gluino the heavier the stop. Therefore we will fix the gluino mass at the electroweak scale as low as possible consistently with the most stringent bounds released by the LHC data~\cite{Aad:2014mra}. So we will fix  $M_3=1.5$ TeV at the low scale. For a fixed value of $\mathcal M$ (after considering the RGE running effects) this will fix the supersymmetry breaking parameter $F$.

We will choose a low value of $\mathcal M$ so that the custodial breaking by the RGE running is minimized. In fact loop corrections to the $\rho$ parameter, that are related to the custodial breaking, are parametrized by $\tan{\alpha_i}-1$, with $\alpha_i=\beta,\theta_0,\theta_1$. Because of the strong effect of the top quark Yukawa coupling, the running differentiates the two soft doublet masses from each other much more than the three triplet ones among themselves. This behaviour which is explicitly shown in Fig.~\ref{fig:running} will result in a much bigger vacuum misalignment in the doublet sector, dictated by the amount of running (i.e.~by the size of the messenger scale $\mathcal M$) and with little dependence on $v_\Delta$. We are therefore left with a situation at the weak scale where $\tan{\beta}\neq 1$ and $(\tan{\theta_0},\tan{\theta_1}\sim 1)$ and so the loop contributions to the $\rho$ parameter coming from the doublet (MSSM) sector will be dominant. As small values of $\mathcal M$ will minimize the resulting value of $\tan{\beta}- 1$ we will fix the messenger scale to $\mathcal M=100$ TeV. In particular as we will see in the next section this will translate, for the benchmark scenario \#1 into $\tan\beta=1.38$ and for the benchmark scenario \#2 into $\tan\beta=1.32$.

\begin{figure}[htb]
\begin{center}
\includegraphics[width=0.48\linewidth]{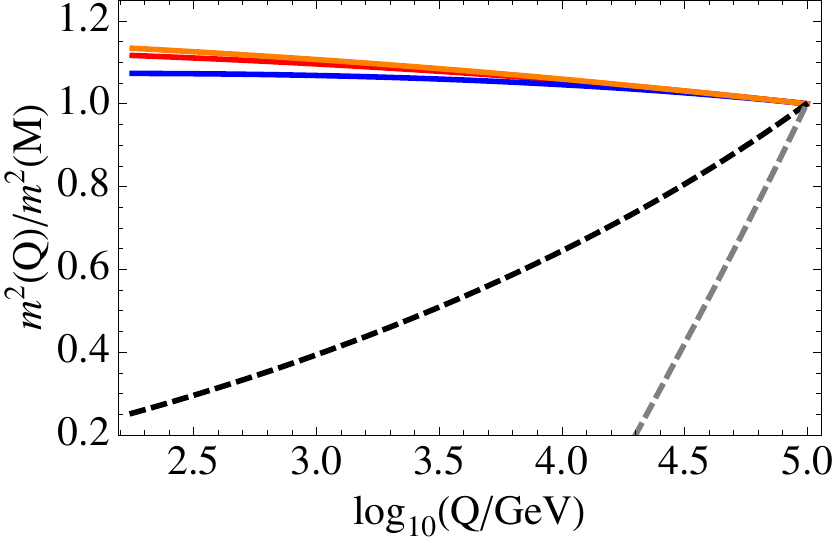}
\includegraphics[width=0.49\linewidth]{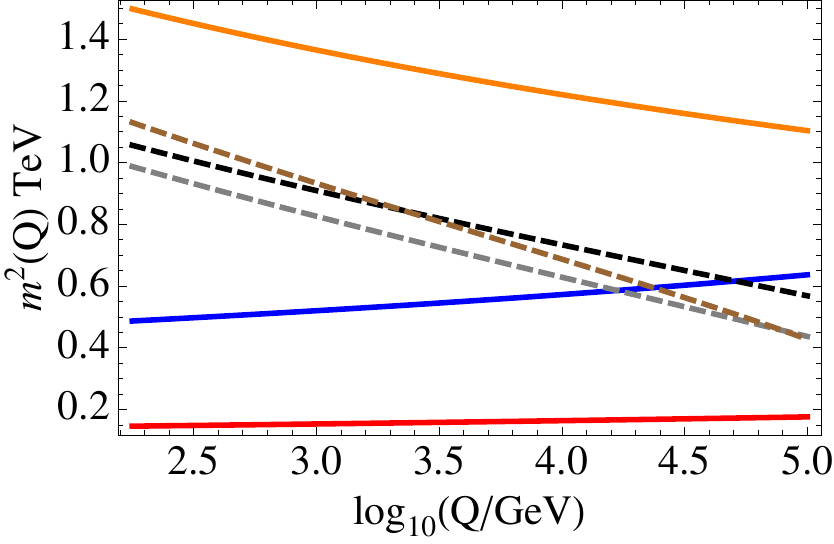}
\caption{\emph{Left panel: Running of  $(m_{H_1}^2,m_{H_2}^2)$ (dashed lines) and $(m_{\Sigma_0}^2,m_{\Sigma_1}^2,m_{\Sigma_{-1}}^2)$ (solid lines), normalized to their values at the messenger scale for benchmark scenario \#1. Right panel: Running of gaugino (solid: $M_3$ orange, $M_2$ blue and $M_1$ red) and squark (dashed: $m_{\tilde{Q}}$ black, $m_{\tilde{t}}$ gray and $m_{\tilde{b}}$ brown) mass parameters for benchmark scenario \#1.}}
\label{fig:running
}
\label{fig:running}
\end{center}
\end{figure}

As a consequence of the low value of the messenger scale the gravitino ($\tilde G$) is the lightest supersymmetric particle (LSP), as usual in gauge mediation. Although the chosen value of $\mathcal M$ is also in agreement with cosmological bounds on the gravitino mass~\cite{Viel:2005qj} the gravitino will not provide the observed relic density by itself, another component will have to enter to fill the DM relic density up to the current observed value. Also, the next to lightest supersymmetric particle (NLSP) will play an important role in the phenomenology of the model. In particular we will see that, in each of the benchmark scenarios studied below, because of the low values of $\sqrt{F}$ the decay $NLSP\to \tilde G+...$ will be prompt, i.e.~it will decay inside the detector but with no displaced vertex, and the experimental signature will be an imbalance in the final state momenta and a pair of photons or charged leptons. 

\subsection{Benchmark scenario \#1: a Bino-like NLSP}
For this scenario we will choose the number of messengers and their couplings with the hidden sector as
\be
\label{eqn:benchmark1}
n_1=1,\, n_3=2, \,n_8=6\quad\text{and}\quad\tilde{\lambda}_1=0.9,\, \tilde{\lambda}_3=0.5, \,\tilde{\lambda}_8=0.1 \,. 
\ee
Note in particular the hierarchy that we establish between $\tilde{\lambda}_8$ and $\tilde{\lambda}_1$. We do this to have as light as possible stops along with sleptons above their experimental bounds. In minimal versions of gauge mediation the contributions given by different gauge groups cannot be disentangled and it is difficult to accommodate light stops without too light sleptons. 

The $SU(2)_L\otimes SU(2)_R$ invariant $\lambda$ of the superpotential will be fixed at the messenger scale such that the correct Higgs mass is reproduced~\footnote{To fit the $125$ GeV value we include the dominant loop contributions to the Higgs mass~\cite{Carena:1995bx}.}, 
\be
\lambda(M)=0.68
\label{eqn:benchmark2}
\ee
We also fix the superpotential parameter $\lambda_3=0.35$, although it will have little effect on the low energy spectrum. The boundary conditions at the messenger scale of $\mu$ (and $\mu_\Delta$) are adjusted to make sure that the vacuum is close enough to the direction $\tan\theta_0=1$, and $\rho$ falls within the allowed $T$ parameter band, $T=0.01\pm 0.12$~\cite{Agashe:2014kda}. In this case we choose both parameters $\mu$ and $\mu_\Delta$ equal at the messenger scale as
\be
\label{eqn:benchmark3}
\mu(\mathcal M)=\mu_\Delta(\mathcal M)= 1.3\,\,\, \text{TeV}
\ee
Of course, the values that will actually fix the Higgs mass are at the EW scale. $\lambda$ and $\mu_\Delta$ are superpotential parameters that we assume to be generated in an $SU(2)_L\otimes SU(2)_R$ invariant fashion.  We show how the running will split these supersymmetric parameters and their EW scale values in Fig.~\ref{fig:running2}.
\begin{figure}[htb]
\begin{center}
\includegraphics[width=0.49\linewidth]{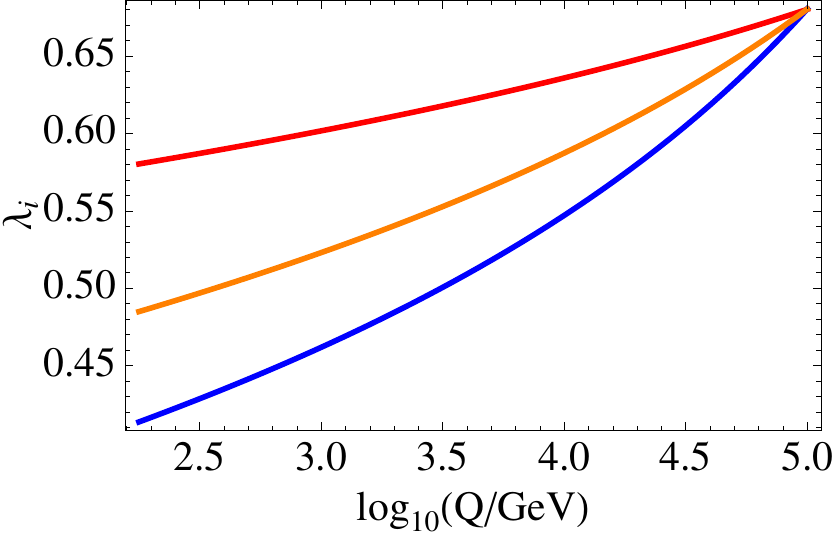}
\includegraphics[width=0.49\linewidth]{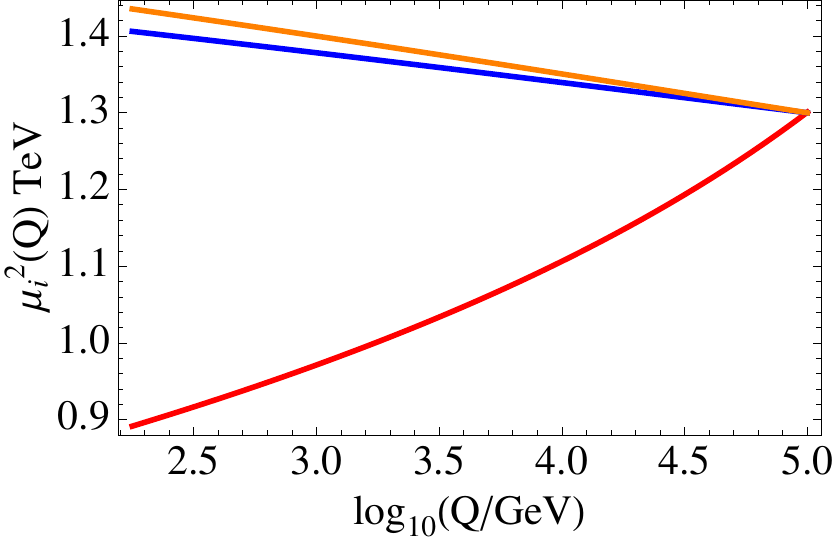}
\caption{\emph{Left panel: For benchmark scenario \#1, running of $\lambda_a$ (red), $\lambda_c$ (orange) and $\lambda_b$ (blue). The $\lambda_{a,b,c}$ F-term contribution to the tree level Higgs is proportional, in the decoupling limit, to~$4\lambda_a^2\cos^4\beta+4\lambda_b^2\sin^4\beta + \lambda_c^2\sin^2 2\beta$. This triplet sector contribution will actually be the only one as the MSSM contribution vanishes when $\tan{\beta}\sim 1$ which is a general feature of our model. Right panel: For benchmark scenario \#1, running of $\mu$ (red), $\mu_{\Delta_a}$ (blue) and $\mu_{\Delta_b}$ (orange).}}
\label{fig:running2}
\end{center}
\end{figure} 

In this scenario the NLSP is a bino-like neutralino that will mainly decay to the gravitino through the following process $\chi_1^0 \rightarrow \gamma \tilde{G}$. If we know its mass and the supersymmetry breaking scale $\sqrt{F}$ we can calculate the average distance travelled in the LAB frame by an NLSP produced with energy $E$ before it decays~\cite{Giudice:1998bp}, 
\be
\label{eq:distance}
L_{\chi_1^0}^{NLSP}=\frac{1}{\kappa_\gamma} \left(\frac{100\,\text{GeV}}{m_{\chi_1^0}}\right) ^5\left(\frac{\sqrt{F}}{100\,\text{TeV}}\right) ^4\sqrt{\frac{E^2}{m^2}-1} \cdot 10^{-2}\, \text{cm},
\ee
with $\kappa_\gamma =|N_{11}\cos{\theta_W}+N_{12}\sin{\theta_W}|^2$, $N_{11}$ and $N_{12}$ being the projections of $\chi_1^0$ to the Bino and Wino respectively (in our case $N_{11}\simeq 1$ and $N_{12}\simeq 0$). In this scenario $\sqrt{F}=94$ TeV and $m_{\chi_1^0}=143$ GeV, this translates in an average distance of flight well below the detector precision ($\sim 0.1$ cm) even if the particle is produced with very high energy and really boosted.

\subsection{Benchmark scenario \#2: $\tilde\tau_R$ as the NLSP}
In this section we present an example of a spectrum where the NLSP is $\tilde{\tau}_R$. We also choose $\mathcal M=100$ TeV and a similar hierarchy between $\tilde{\lambda}$'s, the main difference with \#1 will come in the larger number of messengers,
\be
\label{eqn:benchmarkbis1}
n_1=10,\, n_3=6, \,n_8=5\quad\text{and}\quad\tilde{\lambda}_1=0.9,\, \tilde{\lambda}_3=0.5, \,\tilde{\lambda}_8=0.2 \,. 
\ee
Custodial values in the superpotential are also asjusted at the messenger scale to get the correct Higgs mass and $\rho=1$ at the electroweak scale,
\be
\lambda(\mathcal M)=0.78,\,\lambda_3(\mathcal M)=0.35 \quad\text{and}\quad \mu(\mathcal M)=\mu_\Delta(\mathcal M)= 1.5\,\,\, \text{TeV}.
\label{eqn:benchmarkbis2}
\ee

The $\tilde{\tau}$ will decay into the gravitino through $\tilde{\tau} \rightarrow \tau \tilde{G}$ and we can get its average flight distance from (\ref{eq:distance}) with $\kappa_\gamma =1$. In this case $\sqrt{F} =73$ TeV and $m_{\tilde{\tau}} =343$ GeV and one finds that $L_{\text{\# 2}}^{NLSP} < L_{\text{\# 1}}^{NLSP}$.

\section{Phenomenology of Gauge Mediated SCTM}
\label{sec:pheno}
Figs.~\ref{fig:spectrum} and \ref{fig:spectrumbis} show the spectrum in the two previous benchmark scenarios with light stops, the correct Higgs mass and a non negligible contribution of the triplet sector to EWSB. In particular, in both examples $v_\Delta=25$ GeV, which corresponds to about a $10\%$ of the $W$ and $Z$ masses given by the triplets.
\begin{figure}[htb]
\begin{center}
\vspace{2.1cm}
\includegraphics[width=0.45\linewidth]{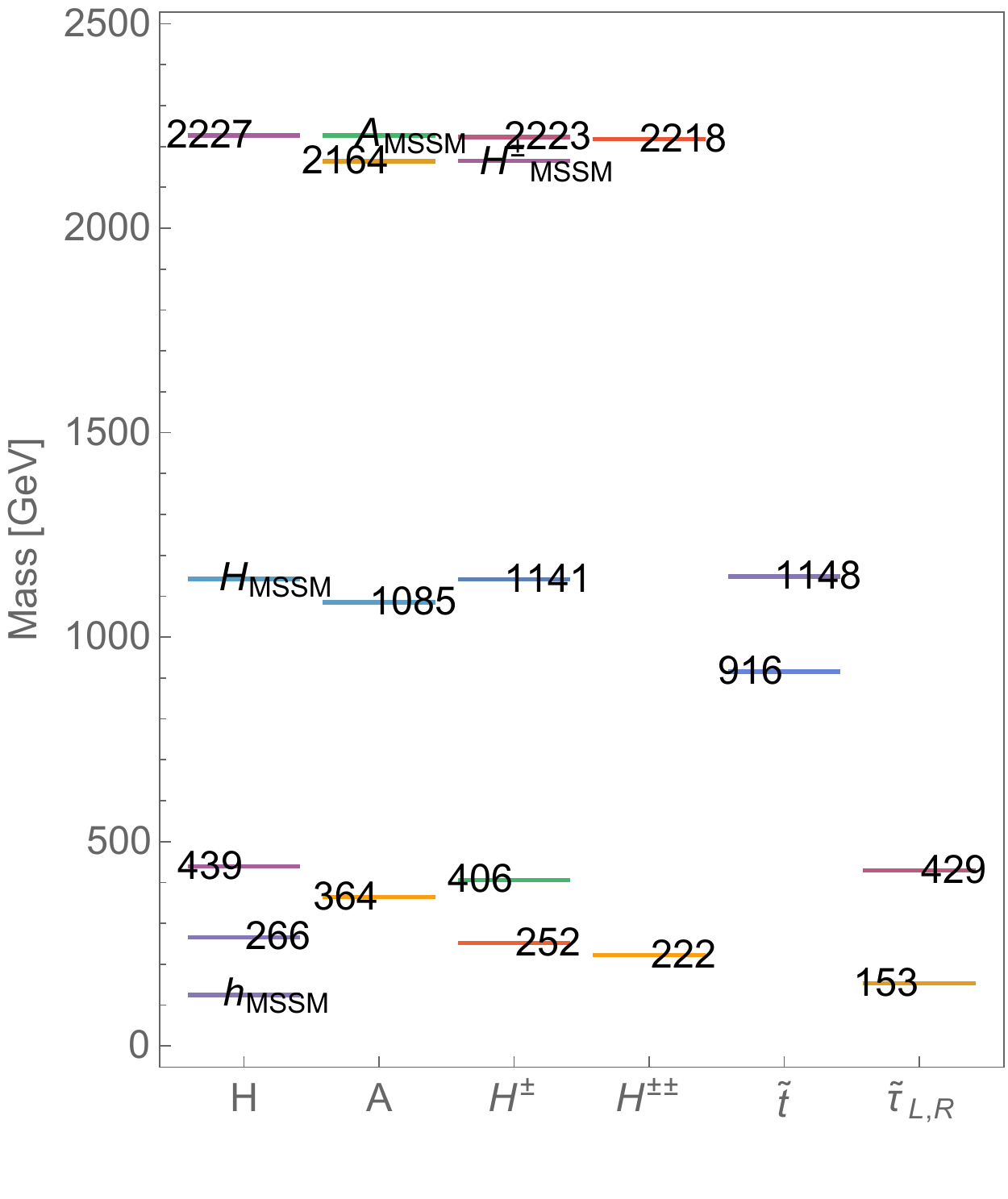}\hspace{5mm}
\includegraphics[width=0.45\linewidth]{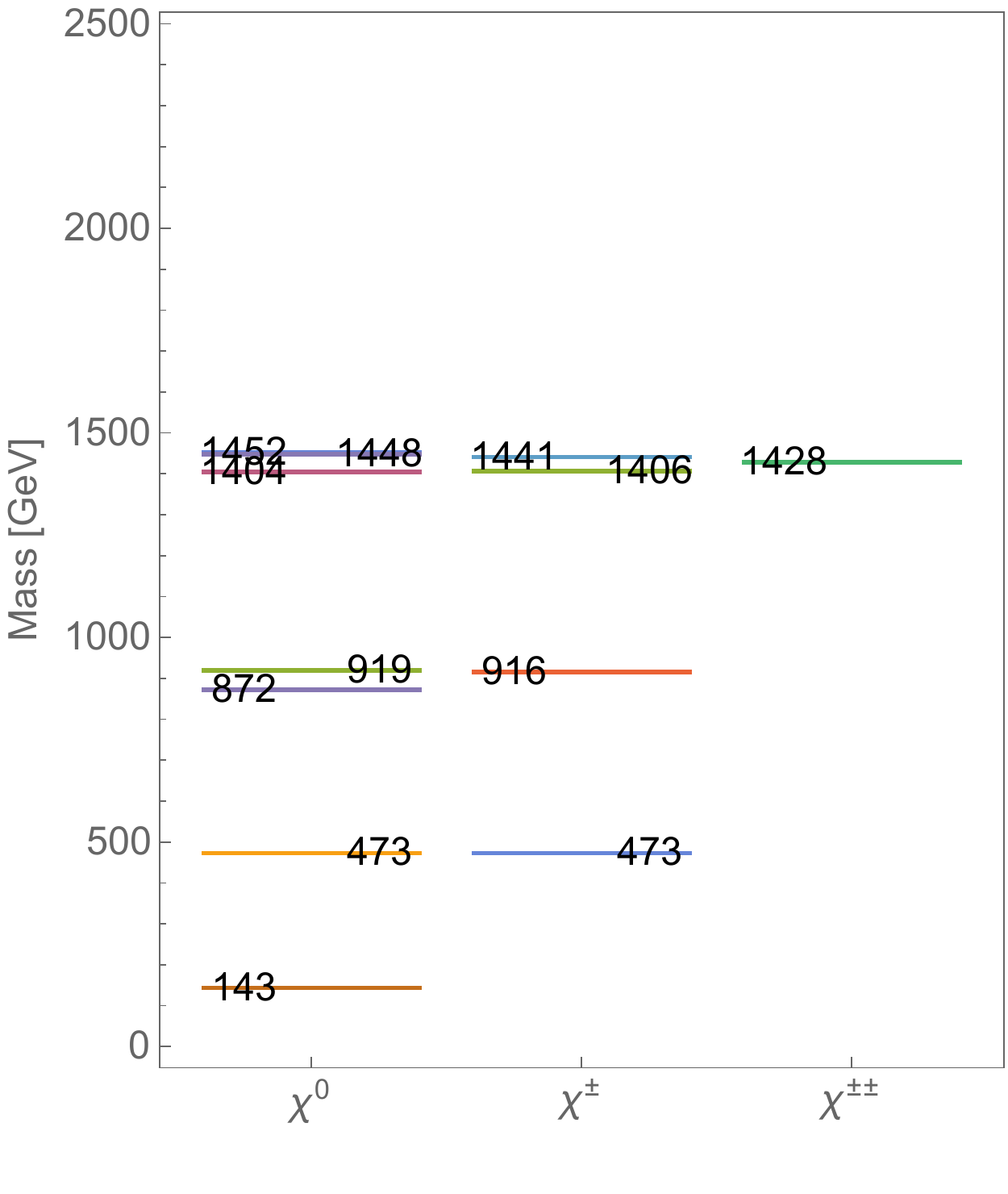}
\caption{\emph{Left panel: Scalar spectrum for scenario \#1. MSSM-like scalars are quoted as so. Right panel: Fermion spectrum for scenario \#1.
}}
\label{fig:spectrum}
\end{center}
\end{figure}
\begin{figure}[htb]
\begin{center}
\vspace{2.1cm}
\includegraphics[width=0.45\linewidth]{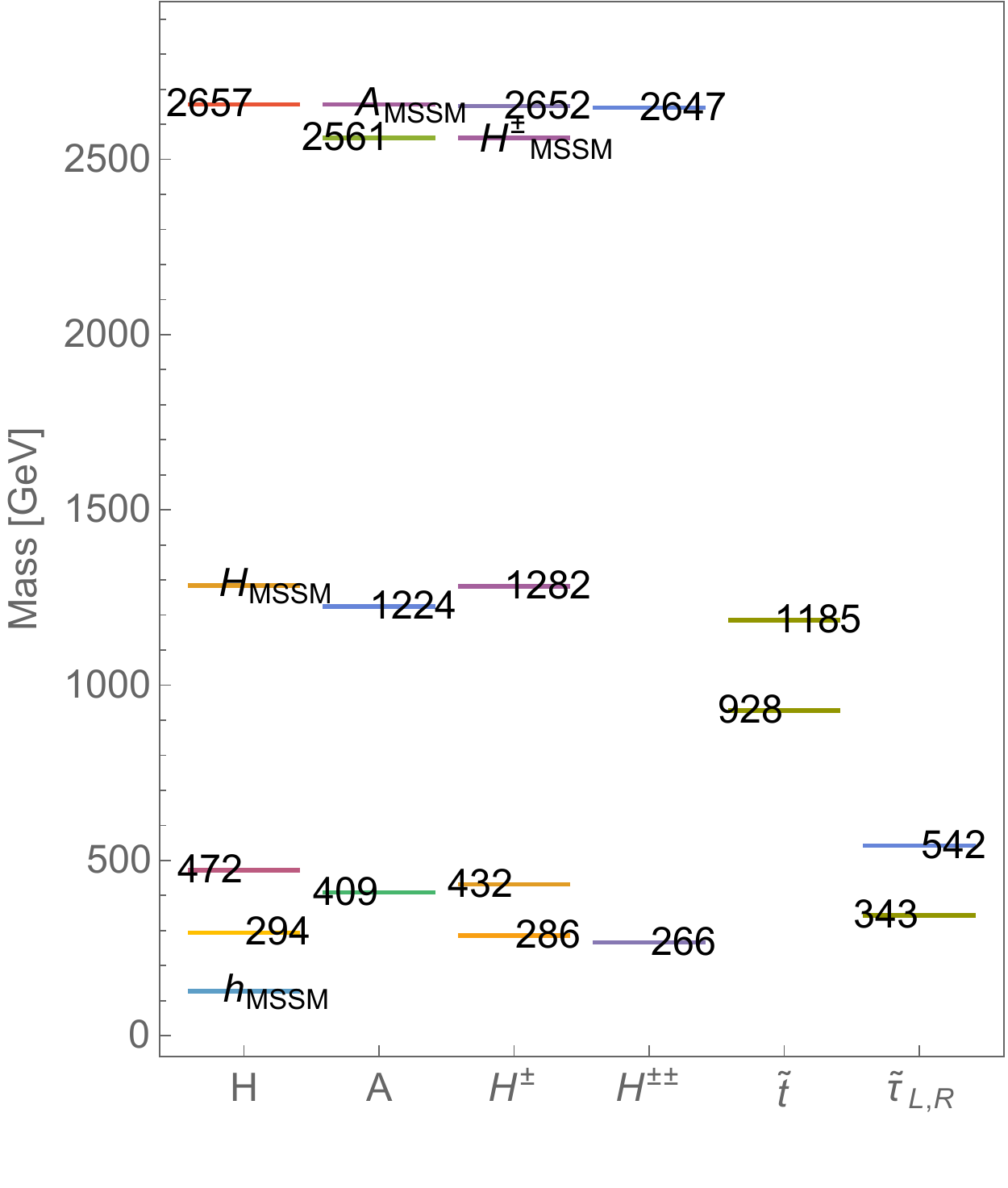}\hspace{5mm}
\includegraphics[width=0.45\linewidth]{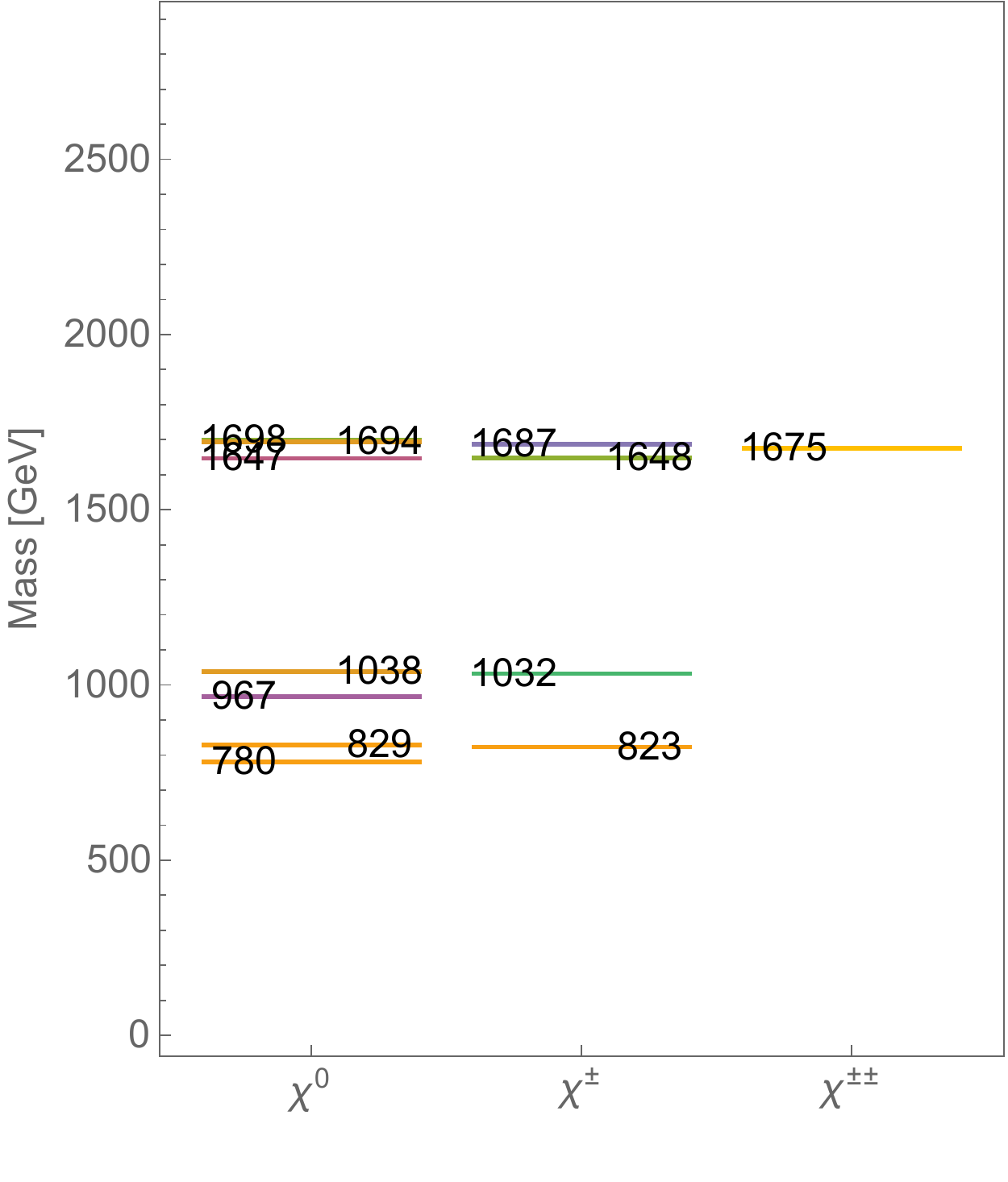}
\caption{\emph{Left panel: Scalar spectrum for scenario \#2. MSSM-like scalars are quoted as so. Right panel: Fermion spectrum for scenario \#2.).
}}
\label{fig:spectrumbis}
\end{center}
\end{figure}
In both scenarios the gravitino cosmology is very simple as $m_{3/2}\sim \mathcal O(\textrm{few})$ eV and the gravitinos are stable particles which do not overclose the Universe as $\Omega_{3/2} h^2\simeq 10^{-3}$. Of course for the same reason we would need a candidate for the dark matter of the Universe, a subject which is beyond the scope of the present paper.

We now look at phenomenological features and possible smoking gun signatures for the present model and in particular for the two benchmark scenarios.

\subsection{Neutralinos and Charginos}
We first analyze the fermionic sector of the theory. The addition of three triplet chiral superfields will enhance the number of neutralinos and charginos. Three extra neutralinos, two new charginos and a doubly charged chargino will be present in the spectrum. Figs.~\ref{fig:spectrum} and \ref{fig:spectrumbis} show the different mass values for scenarios \#1 and  \#2, respectively. As we can see there is a clear hierarchy between states which in part will be determined by the relation (\ref{eq:gauginomass}). In Fig.~\ref{fig:spectrum} this relation is,
\be
\frac{M_1}{\alpha_1}:\frac{M_2}{\alpha_2}:\frac{M_3}{\alpha_3}=1.08:2:1.8\, \quad
[\textrm{scenario \#1}].
\ee
The lightest fermion is the NLSP, a Bino-like neutralino. The next neutralino and first chargino correspond to a Wino-like multiplet, since $M_2$ at the low scale is around $450-500$ GeV. In this scenario $\tilde\chi_2^0$ and the lightest chargino $\tilde\chi^\pm_1$ are (quasi) degenerate in mass. The ATLAS supersymmetric searches~\cite{Aad:2014vma} on $\tilde\chi^0_2\tilde\chi^\pm_1$ production followed by $W$ and $Z$ decays, combined with three-lepton searches, excludes a mass region for degenerate $\tilde\chi_2^0$ and  $\tilde\chi^\pm_1$ between 100 GeV and 410 GeV. These bounds on are satisfied as the mass of $\tilde\chi_2^0$ and  $\tilde\chi^\pm_1$ is $\sim 473$ GeV. The heavier states are doublet-like Higgsinos and tripletinos. 

In scenario \#2 the gaugino mass relation is 
\be
\frac{M_1}{\alpha_1}:\frac{M_2}{\alpha_2}:\frac{M_3}{\alpha_3}=10.8:6:3\, \quad
[\textrm{scenario \#2}].
\ee
and this different hierachy is explicit in Fig.~\ref{fig:spectrumbis}, with a fermion spectrum heavier than in the previous case, also satisfying all present experimental bounds.

\subsection{Sleptons}
ATLAS and CMS searches place strong bounds on slepton masses~\cite{Aad:2014vma,Khachatryan:2014qwa}. These will change depending on whether $\tilde{\tau}_R$ is the NLSP or not. If $\tilde{\tau}_R$ is the NLSP, LHC searches give $m_{\tilde{\tau}_R}\gtrsim 250$ GeV and $m_{\tilde{\tau}_L}\gtrsim 300$ GeV. Bounds are relaxed if we have a neutralino NLSP to which the $\tilde{\tau}_R$ decays. In this case, from the exclusion regions in the  $(m_{\tilde\chi_1^0},m_{\tilde \tau_R})$ plane from decays $\tilde\tau_R\to\tau\tilde\chi_1^0$, it turns out that for $m_{\tilde \chi_1^0}\gtrsim 100$ GeV, there is no LHC constraint on $m_{\tilde\tau_R}$, so that only the LEP bound $m_{\tilde\tau_R}\gtrsim 100$ GeV survives. The latter case applies to our benchmark scenario \#1 where $m_{\tilde \chi_1^0} > 100$ GeV. In the benchmark scenario \#2 we explore the former case and we can see from the mass spectrum that $m_{\tilde{\tau}_R}$ and $m_{\tilde{\tau}_L}$ are above their experimental lower bounds.

\subsection{Higgs scalars}
There are a total of five neutral {\it CP}-even, 4 {\it CP}-odd, 5 singly charged, and two doubly charged massive Higgs scalar fields in this model. With the help of a smooth limit to the MSSM scalar sector, when $v_\Delta \rightarrow 0$, we can identify the MSSM-like states as those which remain light in that limit~\cite{Cort:2013foa}. Due to the small mixing angles between doublets and triplets, the MSSM-like scalars will have a larger doublet component whereas the rest will be mainly composed of triplets. 

Note that the doublet sector is in its decoupling regime and in both cases (Figs.~\ref{fig:spectrum} and \ref{fig:spectrumbis}) there are some light triplet-like scalars. In particular a neutral $H$, a charged $H^\pm$ and a doubly charged $H^{\pm\pm}$ scalar~\footnote{In the custodial case, scalars align themselves under degenerate $SU(2)_V$ multiplets. These light triplet-like scalars correspond to an $SU(2)_V$ fiveplet that, for large enough $v_\Delta$, will be the lightest triplet-like multiplet, just above the Higgs custodial singlet~\cite{Cort:2013foa}. A thorough study of the scalar sector and its departure from the custodial $SU(2)_V$ alignment shows that the degeneracy of the fiveplet masses will be broken in such a way that the hierarchy $m_H>m_{H^\pm}>m_{H^{\pm\pm}}$ is maintained.}. Probing these new triplet-sector states is challenging since the new $SU(2)_L$ triplets do not couple to matter at tree level. For the neutral ones searches for fermiophobic Higgses constrain their masses to be roughly above $194$ GeV~\cite{Chatrchyan:2012vva}. Moreover, the main production process for these states is vector boson fusion and the coupling between a Higgs like scalar and two vector bosons is proportional to its VEV which, for the triplet like states, will be $v_\Delta$, around an order of magnitude smaller than $v$. Due to this, the production cross section will then be smaller than the production of doublet-like scalars and the bound on triplet-like neutral states can be relaxed. 

Although fermiophobic neutral scalars do appear in this model, they are not an exclusive feature of triplet Higgs sectors and cannot be considered a smoking gun of the model. Nevertheless the model has two main characteristic signatures. 
\begin{itemize}
\item
The first one is the appearance of light charged scalars with the coupling $H^\pm W^\mp Z$ and decay channel $H^\pm\rightarrow W^\pm Z$, a decay that is forbidden for charged Higgses coming from doublet representations. This possibility has been explored in~\cite{Aad:2015nfa}. Through the search of $H^\pm\rightarrow W^\pm Z$, and in the context of the non-supersymmetric GM model, ATLAS is able to put bounds on the mass of the triplet-like $H^\pm$. Here we can do a similar consideration to the one we did in searches of fermiophobic scalars. The width of $H^\pm$ is proportional to the squared of $\sin{\theta}=2\sqrt{2}v_\Delta/v$, a factor which parametrizes the amount of mass given by triplets to the $W$ and the $Z$. The experimental bounds grow stronger as $\sin{\theta}\rightarrow 1$ and disappear for $\sin{\theta}<0.5$. In our model $v_\Delta$ is small compared to $v$ so $\sin{\theta}$ is at most $0.35$ and the bounds do not apply. 
\item
The second one is a light doubly charged scalar. Since it does not couple to matter at tree level its only decay mode is $H^{\pm\pm}\rightarrow W^\pm W^\pm$. In~\cite{Kanemura:2014goa} this possibility is studied and bounds on doubly charged scalars are given by looking at possible $H^{\pm\pm}\rightarrow W^\pm W^\pm$ processes. The authors find that with the current LHC data $m_{H^{\pm\pm}}\gtrsim 96$ GeV, a bound obviously satisfied by our benchmark scenarios.
\end{itemize}

Finally, there is also a light pseudoscalar in the spectrum. These are mostly constrained by flavor measurements and electroweak precision observables in the two Higgs doublet model~\cite{Eberhardt:2013uba} and require $m_A\gtrsim 300$ GeV. However, these bounds rely on the fact that the pseudoscalar has to decay primarily on $b\bar{b}$ and $\tau\bar{\tau}$ which happens only when $\tan{\beta}\gg 1$. For our model $\tan{\beta}\sim 1$ at every point of the parameter space so the experimental constraints are relaxed.

\subsection{Higgs couplings}
In this section we explore the properties of the Higgs particle, in particular the normalized couplings of the Higgs to vector bosons and fermions
\be
r_{hXX}=\frac{g_{hXX}}{g_{hXX}^{\rm SM}}\quad
{\rm with}\quad
X=V(W,Z),f(t,b,\tau)
\label{ratios}
\ee
We also look at the loop induced coupling $r_{\gamma\gamma}$ that will contribute to the $h\rightarrow \gamma\gamma$ rate. This rate is dominated in the Standard Model by the propagation of $W$ gauge bosons and top quarks in the loop. The extra contribution from a bosonic or fermionic $Q$-charge sector can be determined from the QED effective Lagrangian~\cite{Shifman:1979eb,Carena:2012xa}
 \be
\mathcal L_{\gamma\gamma}=F^2_{\mu\nu}\frac{\alpha}{16\pi}2 \sum_{J,Q}b_{J}^Q\log\det \mathcal M_{J}^Q(X_R),\quad J=0,1/2;\quad X=H_1^0,H_2^0,\phi^0,\psi^0,\chi^0 
\label{formula}
\ee
where $b_{1/2}^{Q_f}=\frac{4}{3} N_c Q_f^2$ for a $Q_f$-charged Dirac fermion, $b_0^{Q_S}=\frac{1}{3} N_c Q_S^2$ for a complex $Q_S$-charged spin-0 boson ($N_c$ being the number of colors of the corresponding field) and where we have subtracted from the determinant in (\ref{formula}) possible zero-modes (e.g.~charged Goldstone bosons). 

From the values of $r_{hXX}$ one can also compute the predicted signal strength $\mu_{h XX}$ of the decay channel $h\to XX$, with $X=V,\, f,\,\gamma$:
\be
\mu_{h XX}=\frac{\sigma(pp\to h) BR(h\to XX)}{\quad\left[\sigma(pp\to h)BR(h\to XX)\right]_{SM}}~.
\ee
In particular for the gluon-fusion (gF), the associated
production with heavy quarks ($htt$), the associated
production with vector bosons ($Vh$) and the vector boson
fusion (VBF) production processes, one can write
$\mu_{hXX}^{(gF)}=\mu_{hXX}^{(htt)}=
r_{hff}^2 r_{hXX}^2/\mathcal D$ and $\mu_{hXX}^{(VBF)}=\mu_{hXX}^{(Vh)}=r_{hVV}^2 r_{hXX}^2/\mathcal D$. Where 
$\mathcal D \simeq
0.74\, r_{hff}^2 + 0.26\, r_{hVV}^2$.

\begin{table}[h]
\begin{minipage}[c]{0.50\linewidth}
\centering
\begin{tabular}{|c||c|c|c|c|c|}
\hline
Scenario \#1 &${WW}$&${ZZ}$&${b\bar{b}}$&$t\bar{t}$&${\gamma\gamma}$\\ \hline \hline
$r_{hXX}$ &1.05 & 1.04 & 1.01& 1.01&1.22\\ \hline \hline
$\mu_{hXX}^{(gF)},\mu_{hXX}^{(htt)}$&1.07&1.05&1&0.99&1.45\\ \hline
$\mu_{hXX}^{(WF)},\mu_{hXX}^{(Wh)}$&1.16&1.14&1.08&1.07&1.58\\ \hline
$\mu_{hXX}^{(ZF)},\mu_{hXX}^{(Zh)}$&1.14&1.11&1.06&1.05&1.54\\ \hline
\end{tabular}
\par\vspace{0pt}
\end{minipage}
\begin{minipage}[c]{0.45\linewidth}
\begin{tabular}{|c||c|c|c|c|c|}
\hline
Scenario \#2 &${WW}$&${ZZ}$&${b\bar{b}}$&$t\bar{t}$&${\gamma\gamma}$\\ \hline \hline
$r_{hXX}$ &1.05 & 1.04 & 1.01& 1.01&1.18\\ \hline \hline
$\mu_{hXX}^{(gF)},\mu_{hXX}^{(htt)}$&1.07&1.06&0.99&0.95&1.35\\ \hline
$\mu_{hXX}^{(WF)},\mu_{hXX}^{(Wh)}$&1.16&1.15&1.08&1.05&1.46\\ \hline
$\mu_{hXX}^{(ZF)},\mu_{hXX}^{(Zh)}$&1.15&1.14&1.07&1.03&1.45\\ \hline
\end{tabular}
\end{minipage}
\label{tabla}
\caption{\em Left: Higgs couplings and signal strengths for scenario \#1. Right: Higgs couplings and signal strengths for scenario \#2.}
\end{table}%

We show the values of the different couplings and signal strengths for the two benchmark scenarios in Tab.~1. These scenarios are in agreement with the ATLAS current measurements~\cite{atlas:1995} within the present uncertainties. However as the precision will increase, the measurements of Higgs properties will offer one of the most promising avenues to probe this model, in particular through the $r_{h\gamma\gamma}$ coupling. The Higgs is a doublet-like state and therefore its couplings to vector bosons and fermions will not be greatly modified, since the rest of the doublet-like spectrum is heavy enough. However because custodial invariance is broken at the electroweak scale by the RGE running it turns out that there is a corresponding breaking of universality as the parameter $\lambda_{WZ}=r_{WW}/r_{ZZ}$ departs from one. In particular as we can see from Tab.~1, $\lambda_{WZ}-1\simeq 1\%$ for the benchmark scenario \#1 and $\lambda_{WZ}-1\simeq 3\%$ for the benchmark scenario \#2. This breaking of universality was considered in Ref.~\cite{Garcia-Pepin:2014yfa} as one of the possible smoking guns of our model.

Loop induced couplings like $h\gamma\gamma$ can have large modifications. New charged triplet-like light scalar states like $H^\pm$ or $H^{\pm\pm}$ are present and will modify the coupling by circulating along the loop. The lighter these particles are, the greater their effect will be in $r_{h\gamma\gamma}$ and since the masses of triplet-like states scale with $v_\Delta$, $h\rightarrow \gamma\gamma$ will soon put bounds on $v_\Delta$.  

In order to illustrate this point we show in Fig.~\ref{fig:spectrum2} a scenario with the same values of the parameters as the benchmark scenario \#1, but with $v_\Delta=15$ GeV. In this case the scalar spectrum is heavier and the contributions to $r_{h\gamma\gamma}$ are smaller~\footnote{The presence of light charginos could also modify $r_{h\gamma\gamma}$. Note however that in the cases under study $\mu_{a,b}$ is large and no beyond the MSSM light charginos do appear in the spectrum.}.

\begin{figure}[htb]
\begin{minipage}[c]{0.50\linewidth}
\centering
\vspace{8mm}
\includegraphics[width=\linewidth]{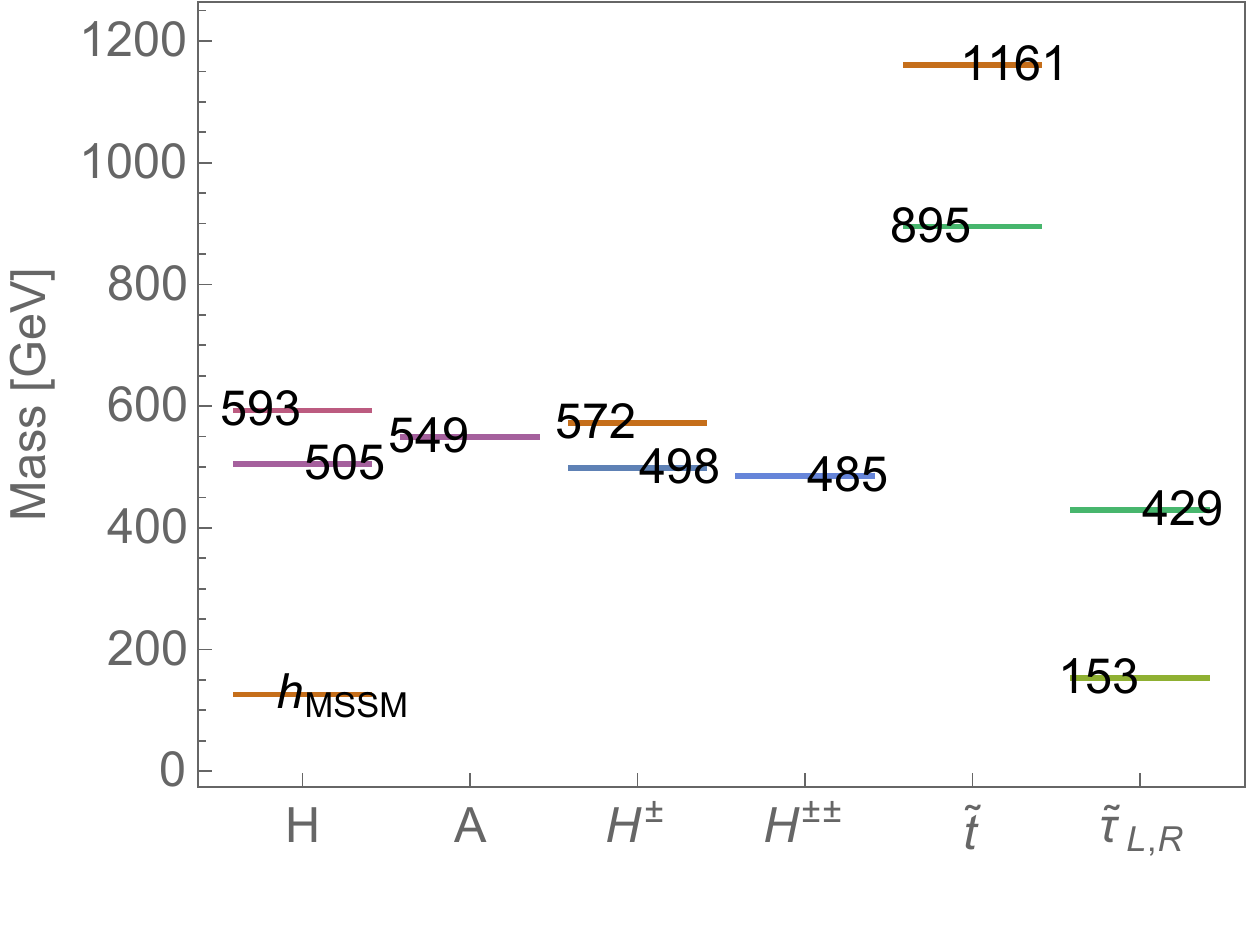}
\par\vspace{0pt}
\end{minipage}
\begin{minipage}[c]{0.45\linewidth}
\centering
\begin{tabular}[c]{|c||c|c|c|c|c|}
\hline
&${WW}$&${ZZ}$&${b\bar{b}}$&$t\bar{t}$&${\gamma\gamma}$\\ \hline \hline
$r_{hXX}$ &1.02 & 1.02 & 1& 1&1.05\\ \hline \hline
$\mu_{hXX}^{(gF)},\mu_{hXX}^{(htt)}$&1.03&1.03&0.98&0.98&1.1\\ \hline
$\mu_{hXX}^{(WF)},\mu_{hXX}^{(Wh)}$&1.07&1.08&1.03&1.03&1.15\\ \hline
$\mu_{hXX}^{(ZF)},\mu_{hXX}^{(Zh)}$&1.08&1.08&1.03&1.03&1.15\\ \hline
\end{tabular}
\par\vspace{0pt}
\end{minipage}
\caption{\emph{Left panel: Scalar spectrum of the benchmark scenario \#1 with $v_\Delta=15$ GeV. $\lambda$, $\mu$ and $\mu_\Delta$ are again adjusted at the messenger scale so the correct Higgs mass is reproduced and $\rho= 1$. Other scalar states are above $1.3$ TeV. Right panel: Higgs couplings and signal strengths with $v_\Delta=15$ GeV.}
}
\label{fig:spectrum2}
\end{figure}

\section{Conclusions}
\label{sec:discussion}

In this paper we have explored the possibility of reconciling the Higgs mass measurement with low scale supersymmetry breaking in the context of gauge mediation. We have done so by implementing a gauge mediated mechanism of supersymmetry breaking to the SCTM, a model where the Higgs sector is extended by three $SU(2)_L$ triplet chiral superfields, whose neutral components can develop large VEVs, which contribute non-negligibly to EWSB consistently with the experimental constraint on the $\rho$ parameter. By adding a non minimal Higgs sector we can generate the correct Higgs mass and still have stops below $1$ TeV. 

In order to satisfy all LHC experimental constraints we have proposed a particular model of general gauge mediation characterized by three species of messengers transforming as non-exotic representations under the SM gauge group, with supersymmetric masses and Yukawa couplings to the spurion field breaking supersymmetry in the hidden sector. In particular we have studied two benchmark scenarios, consistent with all present experimental bounds, with the lightest neutralino (Bino-like) and right-handed stau as NLSP, respectively. For both scenarios the decay of the NLSP is prompt (inside the detector but with no displaced vertex).

We can enumerate a number of characteristic features of our scenarios which depart from the usual minimal gauge mediation in the MSSM:
\begin{itemize}
\item
The first distinct feature is of course (as we already mentioned) that we can reproduce the Higgs mass with light stops ($\sim$ 1 TeV) while in minimal gauge mediation values of the stops mass $\gtrsim$ 5 TeV are required.
\item
There is an extended fermiophobic triplet Higgs sector, absent from the usual supersymmetric extensions of the Standard Model, whose neutral components can acquire a sizeable VEV $v_\Delta$.
\item
The triplet VEVs  can contribute with a non negligible amount to the mechanism of electroweak breaking. A very interesting fact that will be explored by the LHC, as well as the next generation of colliders.
\item
There is a rich phenomenology by new singly ($H^\pm$) and doubly charged ($H^{\pm\pm}$) scalars which, if light enough, can contribute sizeably in loops to $r_{\gamma\gamma}$.
\item
The couplings $H^\pm W^\mp Z$ and $H^{\pm\pm}W^\mp W^\mp$ are proportional to $v_\Delta$ and can thus provide unique signatures for models with extended Higgs sector contributing to the electroweak symmetry breaking mechanism.
\item
The typical pattern for the values of $M_a/\alpha_a$ is strongly spoiled with respect to minimal gauge mediation. Also the sfermion spectrum is completely different from that of typical MGM.
\item
One can measure the amount of custodial breaking by the departure of the universality parameter $\lambda_{WZ}\equiv r_{WW}/r_{ZZ}$ from its custodial value $\lambda_{WZ}=1$.
\end{itemize}

To conclude let us remark that although in this paper we have constructed generic scenarios consistent with all experimental bounds, the constructions are by no means unique. Any of those models should be contrasted with future experimental data, in order to find exclusion regions or some positive signatures which could unveil possible extensions of the Standard Model from the electroweak breaking mechanism. 
\vspace{2cm}


\section*{Acknowledgments}

\noindent We would like to thank Roberto Vega-Morales for useful discussions on the nature of the $H^\mp W^\pm Z$ couplings. This research was supported in part  by the National Science Foundation under Grant No.~PHY-1215979, by the Spanish Consolider-Ingenio 2010 Programme CPAN
(CSD2007-00042), by CICYT-FEDER-FPA2011-25948, by the Severo Ochoa
excellence program of MINECO under the grant SO-2012-0234 and by
Secretaria d'Universitats i Recerca del Departament d'Economia i
Coneixement de la Generalitat de Catalunya under Grant 2014 SGR 1450.


\appendix

\section{Renormalization Group Equations}
\label{sec:rges}
In this appendix we present the complete set of renormalizaton group equations that we have used in our calculations. With $dx/dt=(1/16\pi^2)\beta_x$ we first write the beta functions for the \underline{gauge coupling} constants
\be
\beta_{g_1}  =  \frac{102}{10}g_1^3,\quad
\beta_{g_2}  =  7g_2^3,\quad
\beta_{g_3}  =  -3g_3^3.
\ee
\underline{Yukawa couplings}
\begin{flalign}
\beta_{h_t}&=h_t\left(6h_t^2+h_b^2+6\lambda_b^2+3\lambda_c^2-\frac{16}{3}g_3^2-3g^2-\frac{13}{9}g^{\prime\, 2}\right) \\
\beta_{h_b}&=h_b\left(6h_b^2+h_t^2+6\lambda_a^2+3\lambda_c^2-\frac{16}{3}g_3^2-3g^2-\frac{7}{9}g^{\prime\, 2}\right) \\
\beta_{\lambda_a}&=\lambda_a\left(6\lambda_c^2+14\lambda_a^2+6h_b^2+2\lambda_3^2-7g^2-3g^{\prime\, 2}\right) \\
\beta_{\lambda_b}&=\lambda_b\left(6\lambda_c^2+14\lambda_b^2+6h_t^2+2\lambda_3^2-7g^2-3g^{\prime\, 2}\right) \\
\beta_{\lambda_c}&=\lambda_c\left(8\lambda_c^2+6\lambda_a^2+6\lambda_b^2+3h_t^2+3h_b^2+2\lambda_3^2-7g^2-g^{\prime\, 2}\right) \\
\beta_{\lambda_3}&=\lambda_3\left(6\lambda_3^2+2\lambda_a^2+2\lambda_b^2+2\lambda_c^2-12g^2-4g^{\prime\, 2}\right)\\
\beta_{y_\tau}&=y_\tau\left(  4y_\tau^2+3h_b^2+6\lambda_a^2+3\lambda_c^2-3g_2^2-\frac{9}{5}g_1^2 \right)\ .
\end{flalign}
\underline{Superpotential mass terms}
\begin{flalign}
\beta_{\mu} &= \mu(3h_t^2+3h_b^2+6\lambda_a^2+6\lambda_b^2+6\lambda_c^2-3g^2-g^{\prime\, 2})  \\
\beta_{\mu_{\Delta_a}} &= 2\mu_{\Delta_a}(2\lambda_c^2+2\lambda_3^2-4g^2)  \\
\beta_{\mu_{\Delta_b}} &= \mu_{\Delta_b}(2\lambda_a^2+2\lambda_b^2+4\lambda_3^2-8g^2-4g^{\prime\, 2}) \ .
\end{flalign}
\underline{Gaugino masses}
\be
\beta_{M_1}  =  \frac{102}{5}g_1^2M_1, \quad
\beta_{M_2}  =  14g_2^2M_2, \quad
\beta_{M_3}  =  (-6)g_3^2M_3.
\ee
\underline{Soft scalar mass terms}
\begin{flalign}
\beta_{m_{H_1}^2}&= 2m_{H_1}^2(3h_b^2+6\lambda_a^2+3\lambda_c^2)+6h_b^2(m_Q^2+m_b^2)+12\lambda_a^2(m_{H_1}^2+m_{\Sigma_1}^2) \nonumber \\ &+ 6\lambda_c^2(m_{H_2}^2+m_{\Sigma_0}^2)+6a_b^2+12A_{\lambda_a}^2+6A_{\lambda_c}^2-6g_2^2M_2^2-\frac{6}{5}g_1^2M_1^2-\frac{3}{5}g_1^2S&
\end{flalign}
\begin{flalign}
\beta_{m_{H_2}^2}&=2m_{H_2}^2(3h_t^2+6\lambda_b^2+3\lambda_c^2)+6h_t^2(m_Q^2+m_t^2)+12\lambda_b^2(m_{H_2}^2+m_{\Sigma_{-1}}^2) \nonumber  \\ &+ 6\lambda_c^2(m_{H_1}^2+m_{\Sigma_0}^2)+6a_t^2+12A_{\lambda_b}^2+6A_{\lambda_c}^2-6g_2^2M_2^2-\frac{6}{5}g_1^2M_1^2+\frac{3}{5}g_1^2S &
\end{flalign}
\begin{flalign}
\beta_{m_{\Sigma_0}^2}&= 2m_{\Sigma_0}^2(2\lambda_c^2+2\lambda_3^2)+4\lambda_c^2(m_{H_1}^2+m_{H_2}^2)+4\lambda_3^2(m_{\Sigma_1}^2+m_{\Sigma_{-1}}^2) \nonumber  \\ &+ 4A_{\lambda_c}^2+4A_{\lambda_3}^2-16g_2^2M_2^2 &
\end{flalign}
\begin{flalign}
\beta_{m_{\Sigma_1}^2}&=2m_{\Sigma_1}^2(2\lambda_a^2+2\lambda_3^2)+8\lambda_a^2m_{H_1}^2+4\lambda_3^2(m_{\Sigma_0}^2+m_{\Sigma_{-1}}^2) \nonumber  \\ &+ 4A_{\lambda_a}^2+4A_{\lambda_3}^2-16g_2^2M_2^2-\frac{24}{5}g_1^2M_1^2 +\frac{6}{5}g_1^2S &
\end{flalign}
\begin{flalign}
\beta_{m_{\Sigma_{-1}}^2}&= 2m_{\Sigma_{-1}}^2(2\lambda_b^2+2\lambda_3^2)+8\lambda_b^2m_{H_2}^2
+4\lambda_3^2(m_{\Sigma_0}^2+m_{\Sigma_1}^2)\nonumber  \\ &+ 4A_{\lambda_b}^2+4A_{\lambda_3}^2-16g_2^2M_2^2-\frac{24}{5}g_1^2M_1^2 -\frac{6}{5}g_1^2S  &
\end{flalign}
\begin{flalign}
\beta_{m_Q^2}&= 2m_Q^2(h_t^2+h_b^2)+2h_t^2(m_{H_2}^2+m_t^2)
+2h_b^2(m_{H_1}^2+m_b^2)+2a_t^2+2a_b^2\nonumber  \\ &- 6g_2^2M_2^2 -\frac{32}{3}g_3^2M_3^2-\frac{2}{15}g_1^2M_1^2+\frac{1}{5}g_1^2S &
\end{flalign}
\begin{flalign}
\beta_{m_t^2}&= 2m_t^2(2h_t^2)+4h_t^2(m_{H_2}^2+m_Q^2)+4a_t^2-\frac{32}{3}g_3^2M_3^2-\frac{32}{15}g_1^2M_1^2-\frac{4}{5}g_1^2S &
\end{flalign}
\begin{flalign}
\beta_{m_b^2}&= 2m_b^2(2h_b^2)+4h_b^2(m_{H_1}^2+m_Q^2)+4a_b^2-\frac{32}{3}g_3^2M_3^2-\frac{8}{15}g_1^2M_1^2+\frac{2}{5}g_1^2S&
\end{flalign}
\begin{flalign}
\beta_{m_{\tau_L}^2}&= 2y_\tau^2 m_{H_d}^2+2a_\tau^2+2y_\tau^2m_{\tau_L}^2 +2y_\tau^2m_{\tau_R}^2   -6g_2^2M_2^2-\frac{6}{5}g_1^2M_1^2&
\end{flalign}
\begin{flalign}
\beta_{m_{\tau_R}^2}&= 2(2y_\tau^2m_{H_u}^2+2a_\tau^2+2y_\tau^2m_{\tau_L}^2+2y_\tau^2m_{\tau_R}^2)-\frac{24}{5}g_1^2M_1^2\ .&
\end{flalign}
\underline{Trilinear terms}
\begin{flalign}
\beta_{a_t}
&=
a_t(3h_t^2+6\lambda_b^2+3\lambda_c^2)+h_t(6h_ta_t+12\lambda_bA_{\lambda_b}+6\lambda_cA_{\lambda_c})-\frac{3}{2}g_2^2(a_t-2M_2h_t)\nonumber \\ 
&+ a_t(2h_t^2)+h_t(4h_ta_t)-\frac{8}{15}g_1^2(a_t-2M_1h_t)-\frac{8}{3}g_3^2(a_t-2M_3h_t) \nonumber \\
&+ a_t(h_t^2+h_b^2)+h_t(2h_ta_t+2h_ba_b)-\frac{1}{30}g_1^2(a_t-2M_1h_t)-\frac{3}{2}g_2^2(a_t-2M_2h_t)\nonumber\\ &-\frac{8}{3}g_3^2(a_t-2M_3h_t)-\frac{3}{10}g_1^2(a_t-2M_1h_t) &
\end{flalign}
\begin{flalign}
\beta_{a_b}&=
(a_b(3h_b^2+6\lambda_a^2+3\lambda_c^2)+h_b(6h_ba_b+12\lambda_aA_{\lambda_a}+6\lambda_cA_{\lambda_c})-\frac{3}{2}g_2^2(a_b-2M_2h_b)\nonumber\\ 
&+ a_b(2h_b^2)+h_b(4h_ba_b)-\frac{2}{15}g_1^2(a_b-2M_1h_b)-\frac{8}{3}g_3^2(a_b-2M_3h_b) \nonumber\\
&+ a_b(h_b^2+h_t^2)+h_b(2h_ba_b+2h_ta_t)-\frac{1}{30}g_1^2(a_b-2M_1h_b)-\frac{3}{2}g_2^2(a_b-2M_2h_b)\nonumber \\ &-\frac{8}{3}g_3^2(a_b-2M_3h_b))-\frac{3}{10}g_1^2(a_b-2M_1h_b)&
\end{flalign}
\begin{flalign}
\beta_{A_{\lambda_a}}&=
2(A_{\lambda_a}(3h_b^2+6\lambda_a^2+3\lambda_c^2)+\lambda_a(6h_ba_b+12\lambda_aA_{\lambda_a}+6\lambda_cA_{\lambda_c})-\frac{3}{2}g_2^2(A_{\lambda_a}-2M_2\lambda_a)\nonumber \\ &- \frac{3}{10}g_1^2(A_{\lambda_a}-2M_1\lambda_a))-\frac{6}{5}g_1^2(A_{\lambda_a}-2M_1\lambda_a)\nonumber \\
&+ A_{\lambda_a}(2\lambda_a^2+2\lambda_3^2)+\lambda_a(4A_{\lambda_a}\lambda_a+4A_{\lambda_3}\lambda_3)
-4g_2^2(A_{\lambda_a}-2M_2\lambda_a) &
\end{flalign}
\begin{flalign}
\beta_{A_{\lambda_b}}&=
2(A_{\lambda_b}(3h_t^2+6\lambda_b^2+3\lambda_c^2)+\lambda_b(6h_ta_t+12\lambda_bA_{\lambda_b}+6\lambda_cA_{\lambda_c})-\frac{3}{2}g_2^2(A_{\lambda_b}-2M_2\lambda_b)\nonumber \\ &- \frac{3}{10}g_1^2(A_{\lambda_b}-2M_1\lambda_b))-\frac{6}{5}g_1^2(A_{\lambda_b}-2M_1\lambda_b)\nonumber \\
&+ A_{\lambda_b}(2\lambda_b^2+2\lambda_3^2)+\lambda_b(4A_{\lambda_b}\lambda_b+4A_{\lambda_3}\lambda_3)
-4g_2^2(A_{\lambda_b}-2M_2\lambda_b)-\frac{6}{5}g_1^2(A_{\lambda_b}-2M_1\lambda_b) &
\end{flalign}
\begin{flalign}
\beta_{A_{\lambda_c}}&=
A_{\lambda_c}(3h_t^2+6\lambda_b^2+3\lambda_c^2)+\lambda_c(6h_ta_t
+12\lambda_bA_{\lambda_b}+6\lambda_cA_{\lambda_c})-\frac{3}{2}g_2^2(A_{\lambda_c}-2M_2\lambda_c) \nonumber\\ 
&+ A_{\lambda_c}(3h_b^2+6\lambda_a^2+3\lambda_c^2)+\lambda_c(6h_ba_b+12\lambda_aA_{\lambda_a}
+6\lambda_cA_{\lambda_c})-\frac{3}{2}g_2^2(A_{\lambda_c}-2M_2\lambda_c) \nonumber \\
&+ A_{\lambda_c}(2\lambda_c^2+2\lambda_3^2)
+\lambda_c(4A_{\lambda_c}\lambda_c+4A_{\lambda_3}\lambda_3)-4g_2^2(A_{\lambda_c}
-2M_2\lambda_c)\nonumber \\ &- \frac{3}{10}g_1^2(A_{\lambda_c}-2M_1\lambda_c)&
\end{flalign}
\begin{flalign}
\beta_{A_{\lambda_3}}&=
A_{\lambda_3}(2\lambda_b^2+2\lambda_3^2)+\lambda_3(4\lambda_bA_{\lambda_b}+4\lambda_3A_{\lambda_3})-\frac{6}{5}g_1^2(A_{\lambda_3}-2M_1\lambda_3)-4g_2^2(A_{\lambda_3}-2M_2\lambda_3) \nonumber \\
&+ A_{\lambda_3}(2\lambda_a^2+2\lambda_3^2)+\lambda_3(4\lambda_aA_{\lambda_a}+4\lambda_3A_{\lambda_3})-\frac{6}{5}g_1^2(A_{\lambda_3}-2M_1\lambda_3)-4g_2^2(A_{\lambda_3}-2M_2\lambda_3)\nonumber \\
&+ A_{\lambda_3}(2\lambda_c^2+2\lambda_3^2)+\lambda_3(4\lambda_cA_{\lambda_c}
+4\lambda_3A_{\lambda_3})-4g_2^2(A_{\lambda_3}-2M_2\lambda_3)&
\end{flalign}
\begin{flalign}
\beta_{a_\tau}&=
9y_\tau^2a_\tau+6\lambda_a^2a_\tau+3\lambda_c^2a_\tau+3a_\tau h_b^2+a_\tau y_\tau^2-3g_2^2a_\tau-\frac{9}{5}g_1^2a_\tau\nonumber \\&+y_\tau(12\lambda_aA_{\lambda_a}+2y_\tau a_\tau+6\lambda_cA_{\lambda_c}+6g_2^2M_2+6h_ba_b+\frac{18}{5}g_1^2M_1)&
\end{flalign}
\underline{Soft bilinear terms}
\begin{flalign}
\beta_{m_3^2}&=m_3^2(3h_t^2+3h_b^2+6\lambda_a^2+6\lambda_b^2+6\lambda_c^2-3g_2^2 
-\frac{3}{5}g_1^2)\nonumber \\
&+ \frac{2}{5}\mu(15g_2^2M_2+3g_1^2M_1+30\lambda_cA_{\lambda_c}+15h_ba_b+15h_ta_t+
30\lambda_aA_{\lambda_a}+30\lambda_bA_{\lambda_b}) &
\end{flalign}
\begin{flalign}
\beta_{B_{\Delta_a}}&=4\mu_{\Delta_a}(4g_2^2M_2+2\lambda_cA_{\lambda_c}+2\lambda_3A_{\lambda_3})+2B_{\Delta_a}(2\lambda_c^2+2\lambda_3^2-4g_2^2)) &
\end{flalign}
\begin{flalign}
\beta_{B_{\Delta_b}}&=B_{\Delta_b}(2\lambda_a^2+2\lambda_b^2+4\lambda_3^2-8g_2^2-\frac{12}{5}g_1^2)\nonumber\\ &+\frac{2}{5}\mu_{\Delta_b}(10\lambda_aA_{\lambda_a}
+10\lambda_bA_{\lambda_b}+20\lambda_3A_{\lambda_3}+12g_1^2M_1+40g_2^2M_2)&
\end{flalign}

\bibliography{short}

\providecommand{\href}[2]{#2}\begingroup\raggedright\begin{thebibliography}{10}

\bibitem{Giudice:1998bp}
G.~Giudice and R.~Rattazzi, {\it {Theories with gauge mediated supersymmetry
  breaking}},  {\em Phys.Rept.} {\bf 322} (1999) 419--499,
  [\href{http://xxx.lanl.gov/abs/hep-ph/9801271}{{\tt hep-ph/9801271}}].

\bibitem{Delgado:2013gza}
A.~Delgado, M.~Garcia, and M.~Quiros, {\it {Electroweak and supersymmetry
  breaking from the Higgs boson discovery}},  {\em Phys.Rev.} {\bf D90} (2014),
  no.~1 015016, [\href{http://xxx.lanl.gov/abs/1312.3235}{{\tt
  arXiv:1312.3235}}].

\bibitem{Draper:2011aa}
P.~Draper, P.~Meade, M.~Reece, and D.~Shih, {\it {Implications of a 125 GeV
  Higgs for the MSSM and Low-Scale SUSY Breaking}},  {\em Phys.Rev.} {\bf D85}
  (2012) 095007, [\href{http://xxx.lanl.gov/abs/1112.3068}{{\tt
  arXiv:1112.3068}}].

\bibitem{Draper:2013oza}
P.~Draper, G.~Lee, and C.~E.~M. Wagner, {\it {Precise estimates of the Higgs
  mass in heavy supersymmetry}},  {\em Phys.Rev.} {\bf D89} (2014), no.~5
  055023, [\href{http://xxx.lanl.gov/abs/1312.5743}{{\tt arXiv:1312.5743}}].

\bibitem{Nickel:2015dna}
K.~Nickel and F.~Staub, {\it {Precise determination of the Higgs mass in
  supersymmetric models with vectorlike tops and the impact on naturalness in
  minimal GMSB}},  \href{http://xxx.lanl.gov/abs/1505.0607}{{\tt
  arXiv:1505.0607}}.

\bibitem{Craig:2013wga}
N.~Craig, S.~Knapen, and D.~Shih, {\it {General Messenger Higgs Mediation}},
  {\em JHEP} {\bf 1308} (2013) 118,
  [\href{http://xxx.lanl.gov/abs/1302.2642}{{\tt arXiv:1302.2642}}].

\bibitem{Evans:2013kxa}
J.~A. Evans and D.~Shih, {\it {Surveying Extended GMSB Models with
  $m$$_{h}$=125 GeV}},  {\em JHEP} {\bf 1308} (2013) 093,
  [\href{http://xxx.lanl.gov/abs/1303.0228}{{\tt arXiv:1303.0228}}].

\bibitem{Evans:2015swa}
J.~A. Evans, D.~Shih, and A.~Thalapillil, {\it {Chiral Flavor Violation from
  Extended Gauge Mediation}},  \href{http://xxx.lanl.gov/abs/1504.0093}{{\tt
  arXiv:1504.0093}}.

\bibitem{Byakti:2013ti}
P.~Byakti and T.~S. Ray, {\it {Burgeoning the Higgs mass to 125 GeV through
  messenger-matter interactions in GMSB models}},  {\em JHEP} {\bf 1305} (2013)
  055, [\href{http://xxx.lanl.gov/abs/1301.7605}{{\tt arXiv:1301.7605}}].

\bibitem{Kang:2012ra}
Z.~Kang, T.~Li, T.~Liu, C.~Tong, and J.~M. Yang, {\it {A Heavy SM-like Higgs
  and a Light Stop from Yukawa-Deflected Gauge Mediation}},  {\em Phys.Rev.}
  {\bf D86} (2012) 095020, [\href{http://xxx.lanl.gov/abs/1203.2336}{{\tt
  arXiv:1203.2336}}].

\bibitem{Evans:2011bea}
J.~L. Evans, M.~Ibe, and T.~T. Yanagida, {\it {Relatively Heavy Higgs Boson in
  More Generic Gauge Mediation}},  {\em Phys.Lett.} {\bf B705} (2011) 342--348,
  [\href{http://xxx.lanl.gov/abs/1107.3006}{{\tt arXiv:1107.3006}}].

\bibitem{Calibbi:2013mka}
L.~Calibbi, P.~Paradisi, and R.~Ziegler, {\it {Gauge Mediation beyond Minimal
  Flavor Violation}},  {\em JHEP} {\bf 1306} (2013) 052,
  [\href{http://xxx.lanl.gov/abs/1304.1453}{{\tt arXiv:1304.1453}}].

\bibitem{Calibbi:2014yha}
L.~Calibbi, P.~Paradisi, and R.~Ziegler, {\it {Lepton Flavor Violation in
  Flavored Gauge Mediation}},  {\em Eur.Phys.J.} {\bf C74} (2014), no.~12 3211,
  [\href{http://xxx.lanl.gov/abs/1408.0754}{{\tt arXiv:1408.0754}}].

\bibitem{Chacko:2001km}
Z.~Chacko and E.~Ponton, {\it {Yukawa deflected gauge mediation}},  {\em
  Phys.Rev.} {\bf D66} (2002) 095004,
  [\href{http://xxx.lanl.gov/abs/hep-ph/0112190}{{\tt hep-ph/0112190}}].

\bibitem{Shadmi:2011hs}
Y.~Shadmi and P.~Z. Szabo, {\it {Flavored Gauge-Mediation}},  {\em JHEP} {\bf
  1206} (2012) 124, [\href{http://xxx.lanl.gov/abs/1103.0292}{{\tt
  arXiv:1103.0292}}].

\bibitem{Delgado:2007rz}
A.~Delgado, G.~Giudice, and P.~Slavich, {\it {Dynamical mu term in gauge
  mediation}},  {\em Phys.Lett.} {\bf B653} (2007) 424--433,
  [\href{http://xxx.lanl.gov/abs/0706.3873}{{\tt arXiv:0706.3873}}].

\bibitem{Georgi:1985nv}
H.~Georgi and M.~Machacek, {\it {DOUBLY CHARGED HIGGS BOSONS}},  {\em
  Nucl.Phys.} {\bf B262} (1985) 463.

\bibitem{Cort:2013foa}
L.~Cort, M.~Garcia, and M.~Quiros, {\it {Supersymmetric Custodial Triplets}},
  {\em Phys.Rev.} {\bf D88} (2013), no.~7 075010,
  [\href{http://xxx.lanl.gov/abs/1308.4025}{{\tt arXiv:1308.4025}}].

\bibitem{Garcia-Pepin:2014yfa}
M.~Garcia-Pepin, S.~Gori, M.~Quiros, R.~Vega, R.~Vega-Morales, et~al., {\it
  {Supersymmetric Custodial Higgs Triplets and the Breaking of Universality}},
  {\em Phys.Rev.} {\bf D91} (2015), no.~1 015016,
  [\href{http://xxx.lanl.gov/abs/1409.5737}{{\tt arXiv:1409.5737}}].

\bibitem{Meade:2008wd}
P.~Meade, N.~Seiberg, and D.~Shih, {\it {General Gauge Mediation}},  {\em
  Prog.Theor.Phys.Suppl.} {\bf 177} (2009) 143--158,
  [\href{http://xxx.lanl.gov/abs/0801.3278}{{\tt arXiv:0801.3278}}].

\bibitem{Han:1998pa}
T.~Han, T.~Yanagida, and R.-J. Zhang, {\it {Adjoint messengers and perturbative
  unification at the string scale}},  {\em Phys.Rev.} {\bf D58} (1998) 095011,
  [\href{http://xxx.lanl.gov/abs/hep-ph/9804228}{{\tt hep-ph/9804228}}].

\bibitem{Dvali:1996cu}
G.~Dvali, G.~Giudice, and A.~Pomarol, {\it {The Mu problem in theories with
  gauge mediated supersymmetry breaking}},  {\em Nucl.Phys.} {\bf B478} (1996)
  31--45, [\href{http://xxx.lanl.gov/abs/hep-ph/9603238}{{\tt
  hep-ph/9603238}}].

\bibitem{Dimopoulos:1996ig}
S.~Dimopoulos and G.~Giudice, {\it {Multimessenger theories of gauge mediated
  supersymmetry breaking}},  {\em Phys.Lett.} {\bf B393} (1997) 72--78,
  [\href{http://xxx.lanl.gov/abs/hep-ph/9609344}{{\tt hep-ph/9609344}}].

\bibitem{Dimopoulos:1996gy}
S.~Dimopoulos, G.~Giudice, and A.~Pomarol, {\it {Dark matter in theories of
  gauge mediated supersymmetry breaking}},  {\em Phys.Lett.} {\bf B389} (1996)
  37--42, [\href{http://xxx.lanl.gov/abs/hep-ph/9607225}{{\tt
  hep-ph/9607225}}].

\bibitem{Martin:1996zb}
S.~P. Martin, {\it {Generalized messengers of supersymmetry breaking and the
  sparticle mass spectrum}},  {\em Phys.Rev.} {\bf D55} (1997) 3177--3187,
  [\href{http://xxx.lanl.gov/abs/hep-ph/9608224}{{\tt hep-ph/9608224}}].

\bibitem{Aad:2014mra}
{\bf ATLAS} Collaboration, G.~Aad et~al., {\it {Search for supersymmetry in
  events with large missing transverse momentum, jets, and at least one tau
  lepton in 20 fb$^{-1}$ of $\sqrt{s}=$ 8 TeV proton-proton collision data with
  the ATLAS detector}},  {\em JHEP} {\bf 1409} (2014) 103,
  [\href{http://xxx.lanl.gov/abs/1407.0603}{{\tt arXiv:1407.0603}}].

\bibitem{Viel:2005qj}
M.~Viel, J.~Lesgourgues, M.~G. Haehnelt, S.~Matarrese, and A.~Riotto, {\it
  {Constraining warm dark matter candidates including sterile neutrinos and
  light gravitinos with WMAP and the Lyman-alpha forest}},  {\em Phys.Rev.}
  {\bf D71} (2005) 063534,
  [\href{http://xxx.lanl.gov/abs/astro-ph/0501562}{{\tt astro-ph/0501562}}].

\bibitem{Carena:1995bx}
M.~Carena, J.~Espinosa, M.~Quiros, and C.~Wagner, {\it {Analytical expressions
  for radiatively corrected Higgs masses and couplings in the MSSM}},  {\em
  Phys.Lett.} {\bf B355} (1995) 209--221,
  [\href{http://xxx.lanl.gov/abs/hep-ph/9504316}{{\tt hep-ph/9504316}}].

\bibitem{Agashe:2014kda}
{\bf Particle Data Group} Collaboration, K.~Olive et~al., {\it {Review of
  Particle Physics}},  {\em Chin.Phys.} {\bf C38} (2014) 090001.

\bibitem{Aad:2014vma}
{\bf ATLAS} Collaboration, G.~Aad et~al., {\it {Search for direct production of
  charginos, neutralinos and sleptons in final states with two leptons and
  missing transverse momentum in $pp$ collisions at $\sqrt{s} =$ 8 TeV with the
  ATLAS detector}},  {\em JHEP} {\bf 1405} (2014) 071,
  [\href{http://xxx.lanl.gov/abs/1403.5294}{{\tt arXiv:1403.5294}}].

\bibitem{Khachatryan:2014qwa}
{\bf CMS} Collaboration, V.~Khachatryan et~al., {\it {Searches for electroweak
  production of charginos, neutralinos, and sleptons decaying to leptons and W,
  Z, and Higgs bosons in pp collisions at 8 TeV}},  {\em Eur.Phys.J.} {\bf C74}
  (2014), no.~9 3036, [\href{http://xxx.lanl.gov/abs/1405.7570}{{\tt
  arXiv:1405.7570}}].

\bibitem{Chatrchyan:2012vva}
{\bf CMS} Collaboration, S.~Chatrchyan et~al., {\it {Search for a fermiophobic
  Higgs boson in $pp$ collisions at $\sqrt{s}=7$ TeV}},  {\em JHEP} {\bf 1209}
  (2012) 111, [\href{http://xxx.lanl.gov/abs/1207.1130}{{\tt
  arXiv:1207.1130}}].

\bibitem{Aad:2015nfa}
{\bf ATLAS} Collaboration, G.~Aad et~al., {\it {Search for a Charged Higgs
  Boson Produced in the Vector-boson Fusion Mode with Decay $H^\pm \to W^\pm Z$
  using $pp$ Collisions at $\sqrt{s}=8$ TeV with the ATLAS Experiment}},
  \href{http://xxx.lanl.gov/abs/1503.0423}{{\tt arXiv:1503.0423}}.

\bibitem{Kanemura:2014goa}
S.~Kanemura, M.~Kikuchi, K.~Yagyu, and H.~Yokoya, {\it {Bounds on the mass of
  doubly-charged Higgs bosons in the same-sign diboson decay scenario}},  {\em
  Phys.Rev.} {\bf D90} (2014), no.~11 115018,
  [\href{http://xxx.lanl.gov/abs/1407.6547}{{\tt arXiv:1407.6547}}].

\bibitem{Eberhardt:2013uba}
O.~Eberhardt, U.~Nierste, and M.~Wiebusch, {\it {Status of the
  two-Higgs-doublet model of type II}},  {\em JHEP} {\bf 1307} (2013) 118,
  [\href{http://xxx.lanl.gov/abs/1305.1649}{{\tt arXiv:1305.1649}}].

\bibitem{Shifman:1979eb}
M.~A. Shifman, A.~Vainshtein, M.~Voloshin, and V.~I. Zakharov, {\it {Low-Energy
  Theorems for Higgs Boson Couplings to Photons}},  {\em Sov.J.Nucl.Phys.} {\bf
  30} (1979) 711--716.

\bibitem{Carena:2012xa}
M.~Carena, I.~Low, and C.~E. Wagner, {\it {Implications of a Modified Higgs to
  Diphoton Decay Width}},  {\em JHEP} {\bf 1208} (2012) 060,
  [\href{http://xxx.lanl.gov/abs/1206.1082}{{\tt arXiv:1206.1082}}].

\bibitem{atlas:1995}
{\it {Measurements of the Higgs boson production and decay rates and coupling
  strengths using pp collision data at $\sqrt{s} =$ 7 and 8 TeV in the ATLAS
  experiment}},  Tech. Rep. ATLAS-CONF-2015-007, CERN, Geneva, Mar, 2015.

\end{thebibliography}\endgroup
\bibliographystyle{JHEP}
\end{document}